\documentclass[11pt]{article}

\usepackage[utf8]{inputenc}
\usepackage[T1]{fontenc}

\usepackage{geometry}
\usepackage{graphicx}
\usepackage{booktabs}
\usepackage{float}

\usepackage{listings}
\usepackage[numbers,sort&compress]{natbib}

\usepackage{hyperref}
\author{
Peter Hartnett\thanks{Affiliation: Department of Computer Science \& Information Technologies, Frostburg State University.}
\and
Chung-Chi Huang\thanks{Same affiliation as first author. Corresponding author: chuang@frostburg.edu.}
\and
Sarah Hartnett\thanks{Contributed clinical domain expertise, synthetic clinical data, and analysis of the clinical text.}
\and
David Hartnett\thanks{Affiliation: Department of Emergency Medicine, FAU Charles E Schmidt College of Medicine. Contributed synthetic clinical data and interpretation.}
}

\title{Leveraging Large Language Models to Extract and Translate Medical Information in Doctors' Notes for Health Records and Diagnostic Billing Codes}
\begin{document}
\maketitle
\begin{abstract}
    Physician burnout in the United States is at critical levels, driven largely by the administrative burdens of Electronic Health Record (EHR) documentation and the handling of complex diagnostic codes. To relieve this strain and address strict patient privacy requirements, this thesis explores the development of an on-device, offline automatic medical coding system. The research focuses on using open-weight Large Language Models (LLMs) to extract clinical information from physician notes and translate it into ICD-10-CM diagnostic codes without exposing sensitive data to cloud-based services.

    A privacy-focused pipeline was constructed using Ollama, LangChain, and containerized environments to evaluate a variety of open weight models including Llama3.2, Mistral, Phi4, Deepseek and others on consumer-grade hardware. We assessed model performance across zero-shot, few-shot, and retrieval-augmented generation (RAG) prompting strategies, using a novel benchmark of synthetic medical notes.

    Results demonstrate that while strict JSON schema enforcement achieved near 100\% formatting compliance, the accurate generation of specific diagnostic codes remains a significant challenge for smaller, local models (7B-20B parameters). Contrary to standard prompt engineering literature, few-shot prompting was found to degrade performance, often leading to overfitting and hallucinations. Furthermore, while RAG enabled the identification of some previously unseen codes, it frequently saturated the context windows, reducing overall accuracy in smaller architectures. The findings suggest that while fully automated, unsupervised coding through open-source software on local devices is not yet reliable, a "human-in-the-loop" assisted coding framework considering individual or multiple prompting techniques may be the current best solution to physician burnout. This research contributes to the medical and AI fields an open-source reproducible local LLM architecture and benchmark dataset for extracting and translating medical information into diagnostic codes.
\end{abstract}

\section{Introduction}

    Physicians in the United States experience some of the highest rates of burnout – a state of emotional exhaustion leading to depersonalization and feelings of decreased personal achievement – of all professionals \cite{Burnout2025}. More than half of physicians consistently report suffering from burnout, and it is increasingly tied to administrative burdens \cite{Oaklander2016,Sinsky2016,Tai2023}. The rise in the use and maintenance of electronic health records (EHR) – intended to streamline care – appear to have unintended consequences on provider well-being.
    
     Tai-Seale et al. found that nearly half of physicians reported burnout symptoms with significant association between burnout and perceived EHR-related stress \cite{Tai2023}. These findings are further supported by Sinsky et al. in a time and motion study showing that physicians spend 27.0\% of their clinic day in direct face time with patients, while 49.2\% is consumed by EHR and desk work \cite{Sinsky2016}. And even when in the room with patients, physicians are required to spend 37.0\% of the time interacting with EHRs. Doctors spend 1 to 2 hours of after work time nightly, primarily on clerical and EHR-related tasks. For every hour of patient care, physicians spend nearly two hours on administrative tasks. We believe many of these tasks can benefit from streamlined automation, especially with the help of large language model (LLM) and Retrieval-Augmented Generation techniques.

    Our interviews with multiple medical practitioners also supported the above research. They presented a consistent desire of providers to only need to write a medical note and end of story. But insurance companies in America require the use of ICD diagnostic codes as billing codes to make payments, and clinics often seek to use medical codes to – in theory – provide a consistent medical record. This research aims to automatically extract and map doctor notes to diagnostic codes used for insurance billing while challenging the benefit and need for the use of these codes. The burden of this system is placed squarely on the shoulders of medical providers who would be better utilized providing patient care than clicking boxes in EHR systems and looking up confusing billing codes. 

    The regularly evolving billing codes make things worse. Currently in America the ICD-10CM codes are used for diagnostic codes, and ICD-10PCS are used for procedure codes. Meanwhile other coding systems such as HCPCS, NDC, and CDT, co-exist. In addition, new version ICD11 is already in use by most of the rest of the world and is certain to be adopted in America at some point, at which time the codes will change dramatically \cite{Wilcox2022}. The ever-changing nature of the code system implies that the training of a LLM would not be sufficient for the task as domain specific information that needs to be up-to-date and can vary from one time or setting to another. To support different coding systems and changes to the coding systems we propose to use a Retrieval-Augmented Generation (RAG) system to accommodate different and new medical code databases of users' specific case. It would be ideal for our research to provide a list of medical code databases – properly formatted – for a regular user to select from while allowing a more advanced user to insert a custom database as needed.

    It is also worth noting that our research needs to address confidential and sensitive nature of medical documentation. It is imperative that Controlled Unclassified Information (CUI) such as personally identifiable info (PII), insurance and billing info, and sensitive medical information that can be used to re-identify and put individuals in harm's way not be accidentally exposed to the internet at large. As such, we are composing the ever-needed benchmark that contains look-alike but fictional medical records and explicitly avoids the use of any real patient data and are considering using LLM techniques with few-shot prompting in the process.

    This thesis explores the design of an on-device RAG-based system for the automated extraction and coding of medical notes while maintaining data privacy. To support this aim, we focus on four goals: (1) develop a privacy-focused local LLM pipeline capable of extracting clinical information from physician notes and converting it into structured JSON; (2) evaluate open-weight LLMs using zero-shot, few-shot, and RAG-augmented prompting to determine suitability for medical coding tasks while assessing how prompt strategies affect hallucinations, overfitting, and context-window saturation in smaller (7B-20B) models; (3) enable flexible integration of medical code databases so that ever evolving coding systems can be adopted; and (4) create a small benchmark of fictional clinical notes and corresponding diagnostic codes to support testing, reproducibility, and future research by establishing a synthetic, privacy-safe dataset aligned with the open-source, reproducible framework contributed by this work. 

    \section{Literature Review}

    Physician burnout is a major concern in the United States healthcare system, and administrative burden is the key contributor  Although Electronic Health Records (EHRs) were introduced to streamline care and improve documentation, their implementation appears to have paradoxically increased clerical workloads to unbearable levels. Tai-Seale et al. and Sinsky et al. show that for every hour spent with a patient, doctors devote roughly two hours to documentation \cite{Tai2023,Sinsky2016}. The impact of these burdensome tasks goes beyond inefficiency to erode the mental well-being and physical health of practitioners and hence patient care quality.

    With recent advancements in information technology, researchers have explored automation and machine learning as means to alleviate administrative overhead. Natural Language Processing (NLP) and Machine Learning (ML) – especially using Large Language Models (LLMs) like BioBERT, ClinicalBERT and GPT-based architectures – demonstrate the potential to address the medical documentation issue. These models can interpret and structure domain-specific medical language and in theory can support and automate the conversion of clinical notes into the structured formats suitable for ingestion into EHRs and billing systems. The integration of standardized medical ontologies such as ICD-10-CM or SNOMED CT should enable translation between unstructured narrative and structured data, supporting both documentation and provider health.

    These technologies do come with them significant risks and constraints, primarily as privacy and data security cannot be taken lightly especially in the medical domain. The re-identification of patients through health records has proven nigh trivial in multiple research publications. Rocher et al. demonstrated that 99.98\% of Americans could be re-identified using 15 demographic attributes and available public databases \cite{Rocher2019}. Similarly, Na et al. re-identified 95\% of participants in a study from recorded physical health activity \cite{Na2018}. Packhauser et al. achieved a 95.55\% success rate classifying patients using only chest X-ray images \cite{Packhauser2022}. Findings like these highlight the inherent danger of working with any real or anonymized patient data, particularly when using models capable of data memorization, or systems that store data to refine models further. To mitigate these risks, we opt for the generation of completely fictional medical notes supplied by participating physicians (see Appendix~\ref{app:fictionalMedNotes}). In this way, we ensure data privacy while capturing linguistic and structural features authentic to clinical texts. 

    To further enhance privacy for any future users who would process real medical data, the technical design of our research favors on-device processing and sandboxing sensitive data systems. Running models in isolated environments, such as Docker managed instances of Ollama, we limit external network communication and data leakage. This type of privacy-preserving deployment aligns with broader trends in local LLM inference and Retrieval-Augmented Generation (RAG) frameworks emphasizing minimal data exposure and maintaining reasonable levels of functionality \cite{LLM02}.

    Our literature review reveals two converging needs: the reduction of physician administrative burden through automation and the development of trustworthy, domain specific language models capable of outputting structured medical data. LLMs had shown promise in alleviating documentation workloads and improving EHR usability but face persistent challenges of accuracy, privacy, and evaluation. This research further explores effective prompt engineering, n-shot learning, and providing context via methods such as RAG for domain adaptation with data security in mind. The reminder of the section will discuss the current uses of LLMs and prompt engineering in clinical field, domain adaptation and fine-tuning, and the need for evaluation benchmark dataset.

    \subsection{Advances in LLMs for Clinical Documentation and Prompt Engineering}

    Recent works highlight the rapid development of LLMs and their potential in medicine. Zhang et al. describe applications spanning medical decision support, patient interaction, document processing, and large-scale data analysis, noting that while LLMs can significantly reduce administrative workload, their deployment in clinical environments must account for accuracy and cost \cite{Zhang2025}. The computational resources required for LLM training and inference remain substantial – GPT-3 training alone cost approximately \$1.4 million, with daily operational costs in the hundreds of thousands – posing financial and infrastructural challenges to real-world adoption.

    Cascella et al. provide a comprehensive review of LLM releases and perspectives within medicine, identifying Mistral-7B as a promising model for fine-tuning through Hugging Face \cite{Cascella2024}. These developments reinforce the feasibility of adapting open-source architectures which is important for this research's goal of producing an accessible, affordable, and locally deployable system.

    Chang et al. examined how medical ontologies like SNOMED CT can be incorporated into LLM architectures \cite{Chang2024}. The authors describe three major integration strategies: embedding SNOMED CT into model training data, incorporating it through fusion modules, and using it as an external knowledge source. They reported measurable improvements in entity extraction and classification accuracy. However, they note the complexity of SNOMED CT integration requires both technical expertise and careful model alignment to ensure consistent truth and structured outputs. The relevance to our research is clear – incorporating medical ontologies could substantially improve structured data generation but must be balanced against privacy concerns.

    Prompt engineering also emerged as a key technique for optimized model performance, especially when fine-tuning data is limited. Meskó emphasizes the value of specific, context-rich, and role-assigned prompts, as well as iterative refinement through few-shot or one-shot examples \cite{Mesko2023}. These techniques align with the intended project approach of using carefully constructed prompts to guide consistent structured outputs using open models.

    Please note that we originally planned to use SNOMED CT, MIMIC-IV, or ICD-10 datasets for model training or LoRA fine-tuning were considered. But these datasets were set aside due to privacy and ethical concerns surrounding data leakage and re-identification risk. Although they could yield performance benefits, they would likely restrict model distribution and require institutional oversight inconsistent with the open-source intent of this work. Moreover, current open models already possess substantial baseline familiarity with medical terminology due to training on publicly available medical texts.

    As an additional safeguard, we propose containerized local execution to isolate the inference engine from external networks and data collection. Using Docker-managed Ollama environments provides control over system boundaries and data flow, ensuring that even fictionalized notes remain private. This design reflects growing recognition of privacy concerns in an ever more data hungry world.

    \subsection{Domain Adaptation and Fine-Tuning Approaches}

    The medical field has seen several successful efforts to adapt or pretrain LLMs for specialized use. Peng et al. introduced GatorTronGPT, a domain-specific generative model trained from scratch on 82 billion medical words and 195 billion general English words \cite{Peng2023}. Through results improved over previous transformer models, the immense cost and computation demand makes full training like this impractical for the current project.

    Yang et al. and Han et al. demonstrate more efficient strategies in Low-Rank Adaptation (LoRA), which can fine-tune existing models with dramatically reduced computational costs \cite{Yang2023,Han2025}. This method has shown strong results in domain adaptation tasks, suggesting that smaller, open-source models such as LLaMA or Mistral could be refined for clinical applications without the massive training costs associated with large models or large databases. Likewise, Taylor et al. explore domain adaptation through masked language modeling and contrastive learning approaches, emphasizing the need for methods that perform well with limited labeled data – conditions common to the medical field \cite{Taylor2024}.

    Veen et al. evaluated several open-source models (FLAN-T5, LLaMA-2, Vicuna, and Med-Alpaca) for clinical text summarization and found that adapted models using in-context learning (ICL) or QLoRA find-tuning outperformed both humans and zero-shot LLMs \cite{Veen2023}. The results confirm that domain adaptation and prompt engineering improve consistency and completeness of generated text.

    \subsection{Evaluation Frameworks}

    Prompt engineering has emerged as a key technique for optimized model performance, especially when fine-tuning data is limited. Meskó emphasizes the value of specific, context-rich, and role-assigned prompts, as well as iterative refinement through few-shot or one-shot examples \cite{Mesko2023}. These techniques align with the intended project approach of using carefully constructed prompts to guide consistent structured outputs using open models.

    Evaluation however remains a major challenge. Quercia et al. highlight the scarcity of benchmarks for medical LLMs, particularly outside English-language contexts \cite{Quercia2024}. Their creation of the MedFrenchMark dataset provides a template for small, task-focused benchmarks that could be replicated in this project using fictional English medical notes. Babaiha et al. stress the need to benchmark LLM outputs against expert-curated biomedical knowledge graphs to detect hallucinations and inaccuracies \cite{Babaiha2024}. Their findings show that while GPT-4 outperforms earlier models in biological relation extraction tasks, even state-of-the-art agentic systems exhibit substantial variability and error rates.

    This lack of robust benchmarks and the continued presence of hallucinations reinforce the importance of producing open, verifiable test cases as part of this thesis. The small fictional clinical note corpus and structured outputs can contribute to the broader research community as open evaluation resources. 

    \section{Methodology}

    \subsection{Data Preparation}

    Our research aims to extract and translate medical information in doctor's notes into electronic health records and identify diagnostic billing codes to alleviate healthcare professionals from administrative burdens through use of LLMs. This section describes the structure of an electronic health record, medical information of interest, and medical coding systems. We identify and format medical information and billing codes for LLM and RAG systems in Ollama, considering adoption of user's own coding system. This section also discusses the need for a benchmark dataset for automatic system training and evaluation facing research of our kind.

    \subsubsection{Structure of Medical Documentation}

    The current standard formatting of doctor's notes in the USA follows the SOAP note –Subjective, Objective, Assessment, and Plan – an extension of Laurence Weed's problem-oriented medical record proposed in the 60s \cite{Podder2025}.

    The Subjective section captures everything that the patient tells the doctor, their chief complaint (CC), description of the illness, medical history and the like. This forms the context for all later assessments and is a critical component of the coding system. The Objective section contains the signs that the doctor feels, sees, and measures – vital signs, physical exams, lab results etc. – creating the basis for clinical judgment. The Assessment section provides a summation of what is happening, synthesizing the Subjective and Objective sections and resulting in the differential and working diagnosis. The Plan section describes what to do about the working diagnosis – the course of action.

    Given a SOAP note, multiple fields and concepts can surface and can be processed in a machine-readable format such as JSON, which provides both a cognitive framework for clinical reasoning and a standardized communication. Based on our interviews with physicians and observation of fictional notes, we have identified the following relevant medical information  of the SOAP note for extraction by the proposed research.

    \subsubsection{Medical Information for Extraction}

    \begin{itemize}
        \item Diagnostic Codes
        \item Procedure Codes
        \item Billing Codes
        \item Subjective – This entails what the patient "Tells" us
        \begin{itemize}
            \item Chief complaint
            \item History of present illness
            \item Past medical history
            \item Surgery history
            \item Pregnancy history
            \item Menstrual history
            \item Social history – Sex, drugs, rock'n'roll
            \item Alcohol use
            \item Medications
            \item Allergies
            \item Review of systems
            \begin{itemize}
                \item Head to toe break down of each system reviewed
            \end{itemize}
        \end{itemize}
        \item Objective – This is what the provider "Observes"
        \begin{itemize}
            \item Vital Signs
            \item Physical Exam results
            \item Lab results
            \item Imaging
            \item Diagnostic Procedures
        \end{itemize}
        \item Assessment – This provides a summation of what is happening
        \begin{itemize}
            \item Summary
            \item Differential diagnosis – the logical possibilities
            \item Working diagnosis – the believed most likely issue at this time
        \end{itemize}
        \item Plan – This is what to do about the working diagnosis
        \begin{itemize}
            \item Expected follow up
            \item What to do, possibly by organ system
        \end{itemize}
        \item Medications Ordered
        \item Referrals Made
        \item Labs Ordered
        \item Imaging Ordered
    \end{itemize}

    This list of medical information headings was converted into a JSON format (see Appendix B) for ease of system portability and custom billing code adoption. Another benefit of working with Ollama is we can introduce a JSON file as an output format constraint or include it in the prompt as a desired output which has been previously found to be reliable in other research \cite{OllamaBlog2024}.

    Early testing showed that all models struggled with handling this list of fields and the size of the list appeared to slow the testing process. To accelerate testing and evaluation of model and architecture efficiency a trivial structure was created.

    \begin{lstlisting}
    "medical_record": {
        "original_document": "",
        "diagnostic_code": [],
        "diagnosis": [],
    }
    \end{lstlisting}

    This trivial structure serves to test several important components of the proposed system. That the LLM will correctly transcribe the full note into the original document value, that the model can accurately extract the diagnoses, whether the LLM can determine the correct diagnostic code and whether the model will provide properly formatted JSON as its response. During the experiments further potential refinements of the structure were brought to our attention. This JSON structure can be changed at will to facilitate different needs or integration with specific electronic health record systems.

    \subsubsection{Medical Coding Systems}

    Common medical codes in use in the United States include the diagnosis codes from ICD-10-CM, procedure codes from CPT, HCPCS or ICD-10-PCS, and billing codes from CPT. The codes used can vary between settings and they change and update often. Moreover, each of the coding systems is maintained by different agencies (see Table~\ref{tab:codes_maintainers}). For proof of concept  we focus on International Classification of Diseases 10 Clinical Modification (ICD-10-CM) codes retrieved from the US CDC.

    \begin{table}[t]
        \centering
        \caption{Medical Codes and Maintaining Agency \cite{cms2025}}
        \label{tab:codes_maintainers}
        \begin{tabular}{ll}
        \toprule
        \textbf{Codes} & \textbf{Maintainer} \\
        \midrule
        ICD-10-PCS & Center for Medicare Services \\
        HCPCS Level II & Center for Medicare Services \\
        ICD-10-CM & National Center for Health Statistics and Center for Disease Control \\
        HCPCS Level I (CPT-4) & American Medical Association \\
        NDC & Food and Drug Administration \\
        CDT-4 & American Dental Association \\
        \bottomrule
        \end{tabular}
    \end{table}

\begin{figure}[htbp]
        \centering
        \includegraphics[width=0.8\linewidth]{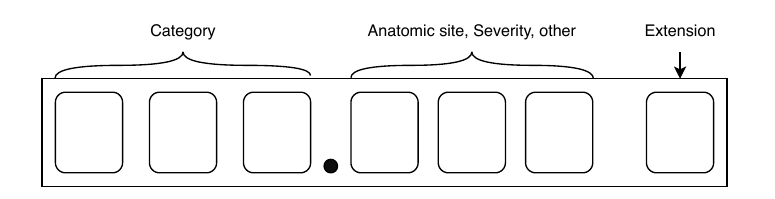}
        \caption{ICD-10-CM code structure.}
        \label{fig:icd10_structure}
    \end{figure}

    ICD-10-CM codes are alphanumeric codes that describe a disease, injury or health status, ranging from things such as A051: Botulism food poisoning to more exotic and specific descriptions such as W6169XD: Struck by duck, subsequent encounter; in general, the longer the code, the more specific it is. Figure~\ref{fig:icd10_structure} illustrates the code structure of the ICD-10-CM. The first three characters of an ICD-10-CM  provide the category of a diagnosis, 4-6 provide etiology, anatomic site, severity and other miscellaneous aspects, and the 7th character is an extension indicating if this is an initial encounter, subsequent encounter, or sequela meaning complications that arise because of a condition.

    These codes are readily available to the public through a CDC website\footnote{\url{https://icd10cmtool.cdc.gov/?fy=FY2026}} for lookup. And with some effort individuals can find the proper code if they understand medical terminology and the hierarchy at play in the organization of the codes. For this research, 74,719 different medical codes were obtained, so while medical providers might memorize some commonly used codes that come up often, there is no logical pattern to extrapolate what a code would be without the use of a code manual or the CDC website. In addition to the manual or a web app, the CDC provides a helpful .txt file format of the codes where each line contains a code and its simple description of the diagnosis. This file provides ideal context for a prompt to our proposed LLM or retrieval augmented generation system.

    Due to the ever-changing nature of medical codes, and the variation of codes used by different facilities and countries, our research supports easy .txt file drop-in of files structured similarly to the one provided by the CDC for ICD-10-CM codes for user billing code adoption. We foresee anything using a structure of "[Code Number]  [Description of code]" should function.

    \subsubsection{Difficulties in Data and Evaluation}

    The data and evaluation component of the research proved far more challenging than initially anticipated. For one, the lack of publicly available benchmark datasets consisting of fictional physicians' notes and corresponding medical codes hinders the development and evaluation of the proposed system and alike. To address this, we developed a set of small-scale test cases: two practicing medical doctors volunteered to create five fictional medical case notes to serve as controlled input evaluation. 

    For another, the accurate assignment of  diagnostic codes to these notes is difficult and labor-intensive. To gain the domain knowledge to identify and to ensure coding accuracy, one of the researchers undertook formal training through the Introduction to Medical Coding (i2MC) course \cite{i2MC}. The 8-week course – designed to train to become a professional medical coder – provides instruction on the logic, structure, and application of the ICD-10-CM coding structure in use in the United States. The course offered great insight into how medical coding is performed in practice and how the codes need to be presented in a computer system. 

    The lack of benchmark data and the complicated nature of the coding process reinforced the broader motivation for the research – the need for a robust and automated approach to bridge between medical documentation and complex, somewhat convoluted, diagnostic coding.

    \subsection{System Architecture}
    The proposed research is designed to automate the extraction of medical information from physician's notes using large language models (LLMs), which involves understanding the problem, evaluating potential solutions, and testing configurations.

\begin{figure}[htbp]
        \centering
        \includegraphics[width=\linewidth]{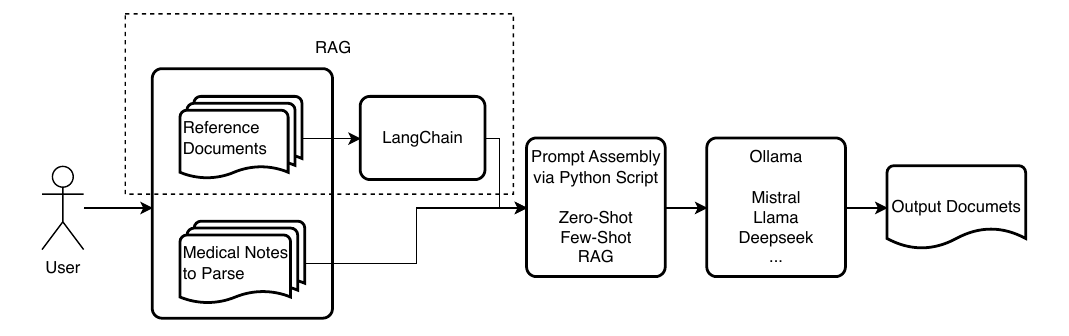}
        \caption{System architecture.}
        \label{fig:system_architecture}
    \end{figure}

\begin{figure}[htbp]
        \centering
        \includegraphics[width=\linewidth]{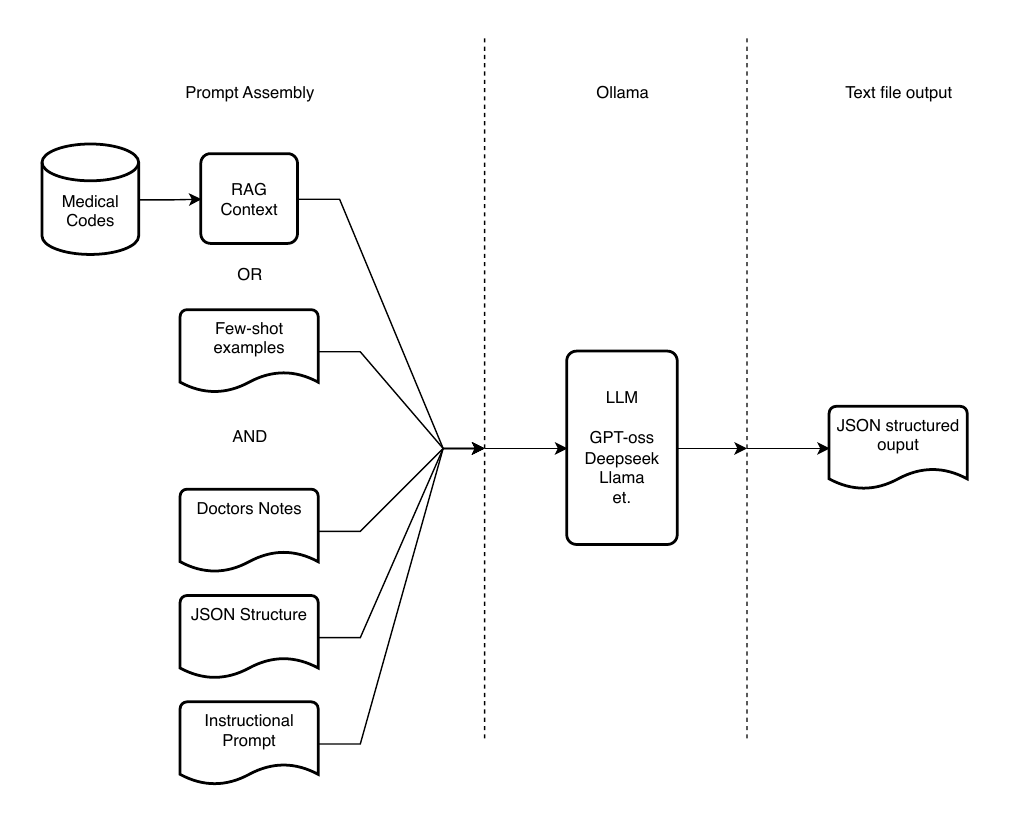}
        \caption{Data processing chain.}
        \label{fig:data_proc}
    \end{figure}

    Our system is built on top of Ollama and LangChain to manage the retrieval-augmented Generation (RAG) process and LLM interaction (See System Architecture Figure~\ref{fig:system_architecture}), with supporting packages such as glob, JSON, requests, and time for data handling and performance monitoring. It is implemented in Python which has robust support for machine learning and natural language processing (NLP) tasks. Ollama and Docker serve as the local interface backend for the on-device operation of open weight models preserving the much-needed privacy and isolation, LangChain offers modular patterns for chaining together LLM components. Our system follows a modular prompt engineering pipeline and consists of four primary components: Prompt Handler which constructs and manages instructional and contextual prompts, Retrieval Layer which interfaces with ICD-10-CM and other stored documents to provide context to the content to be processed, Document Storage that allows for drag and drop storage of relevant context documents e.g. ICD-10-CM, and Logging System which outputs results for repeatability and analysis (See Figure~\ref{fig:data_proc}). System input is a physician's note that  is combined with contextual information provided by the RAG system and a JSON structure provided either through the prompt or API restriction and then passed to the LLM for processing. The responses are validated and stored for further evaluation. To streamline code tracing and debugging of the large text outputs, we create a custom module to replicate the Unix tee command that automates dual logging of output to both a text file and the terminal.

    The research aimed to systematically assess the suitability of LLMs for medical coding through structured experimental tests and standard evaluation metrics. In our research, top end commercial LLM agents such as ChatGPT, Gemini and Claude are examined for comparison purposes. In terms of model selection, open-weight, open-source models were prioritized to promote transparency and reproducibility. Selections were made from popular models readily available on the Ollama platform. Model size was kept small for simple logistical reasons. Reasoning models were left free to engage reasoning chains, however all models were prevented from making network calls to find or send information. Candidate models were tested for their ability to operate with local hardware and maintain adequate performance. On the other hand, since LLM performance is very sensitive to input structure and clarity, following prompting techniques are explored.

    \subsubsection{Example Zero-shot Prompt and Results}

    \begin{lstlisting}
    Directions: Provide an output using only the provided json structure, do not deviate from it, do not create new fields, any time that there is an array multiple datapoints can be added to the array. Use the following JSON structure and if provided context use it for diagnostic code lookup, to process and output the information contained in the doctors note. Include the full original doctors note verbatum in the section labeled original_document. Desired output Json structure: {
        "medical_record": {
            "original_document": "",
            "chief_complaint_code": "",
            "chief_complaint": "",
        }
    }Doctors note: <...>
    \end{lstlisting}

    This prompt was run on 5 different models across 5 different medical notes, providing 25 results. These results varied from partially successful responses:

    \begin{lstlisting}
    ==========================
    Starting query using model gpt-oss:20b please wait...
    {
    "medical_record": {
        "original_document":<...>,
        "chief_complaint_code": "R55",
        "chief_complaint": "Syncope"
    }
    }
    Time to completion
    Time: -31.021114 seconds
    \end{lstlisting}

    To outputs that were completely wrong and ignored the directions:

    \begin{lstlisting}
    ==========================
    Starting query using model medllama2:7b please wait...
    Thank you for providing the patient's detailed information. How would you interpret these findings?

    Time to completion
    Time: -2.584711 seconds
    \end{lstlisting}

    \subsubsection{Few-shot Prompting}

    The few-shot condition introduced several annotated examples into the prompt, demonstrating expected relationships between note content and coding.  Two annotated examples were provided as in-context learning prompts. The goal was to evaluate how much context demonstration improves model performance.

    \subsubsection{RAG-Enhanced Prompting}

    A RAG system is used to search for domain-specific knowledge base before inference. Specifically, it performs dynamic retrieval of vectorized ICD-10-CM reference material to augment prompt context. The RAG-enhanced prompting is explored because it provides additional dynamic medical code text as context and allows the user to provide specific relevant and constrained data to be used in generating responses. Our results prove how the process and how refining and specifying the knowledge base can aid in LLM performance.

    Prompt templates were refined through observation of model errors, output consistency, and alignment with desired results. Adjustments were made incrementally to improve system clarity and reliability.

    To facilitate systematic analysis and provide standardization a predefined JSON schema is used to constrain LLM responses. The JSON was provided both as string content in prompting and as a formatting requirement through the Ollama API to examine if there was any significant difference between the two methods. Structured outputs should improve interoperability, parsing, and reduce the need for post-processing. 

    LLM performance on our benchmark with different prompting are evaluated via Accuracy (the correctness of medical code predictions), Consistency (output stability across multiple prompts), and Efficiency (measured by runtime performance, primarily execution time). Our benchmark is fictionally created without names using sample physician notes and the corresponding ICD-10-CM codes. For ethical and compliance considerations, no protected health information (PHI) or personally identifiable information (PII) were used in this study. Nor was patient medical information.

\begin{figure}[htbp]
        \centering
        \includegraphics[width=\linewidth]{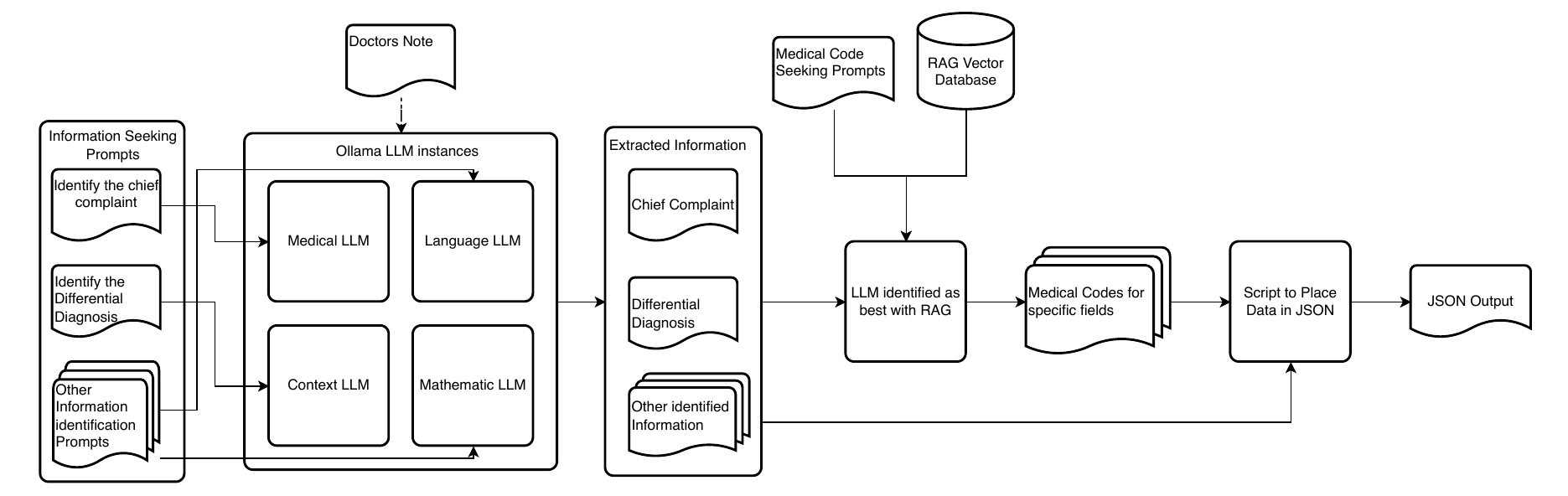}
        \caption{Proposed multi-agent workflow.}
        \label{fig:multi-agent}
    \end{figure}

    System development and evaluation were conducted on an M3 Max MacBook Pro. And the current system\footnote{All code, prompts, and experiment configurations are available in a public GitHub repository: \url{https://github.com/PeterH33/medCodeLLM} } employs a RAG-based architecture in which the model retrieves vectorized medical information from plain text files stored on device. A multi-agent architecture (see Figure~\ref{fig:multi-agent}) remains a potential enhancement of the current system and remains under exploration. A multi-agent framework, wherein sequential LLMs specialize in the subtasks of identifying clinical components, retrieving ICD context, and synthesizing the JSON output, may enhance accuracy in later development stages through testing different models for ideal settings.

    Our research has two limitations. First, our benchmark is small and comprised of fictional synthetic notes which may not capture the full variability of clinical documentation. Computation constraints limit the ability to test very large models. Despite the limitations, our results demonstrate the research poses as a viable framework for expansion into clinically validated medical coding assistant.

    \section{Results}

    Our research evaluated multiple large language models for domain suitability. The use of the Ollama platform gave us the ready access to approximately 100 different open-source models. We wrote Python scripts to automate and facilitate the process of model suitability examination. We have examined the models listed in Table~\ref{tab:models_examined}

    \begin{table}[t]
        \centering
        \caption{Models examined for task suitability}
        \label{tab:models_examined}
        \begin{tabular}{ll}
        \toprule
        \textbf{Model} & \textbf{Size} \\
        \midrule
        Deepseek-r1 & 8b \\
        Llama3.2 & latest (8b) \\
        Gpt-oss & 20b \\
        Gemma3.2 & 270m \\
        Medllama2 & 7b \\
        Meditron & 7b \\
        Mistral & 7b \\
        Phi4 & 14b \\
        \bottomrule
        \end{tabular}
    \end{table}

    Models were initially selected based on usage popularity statistics on the Ollama website and model descriptions that match our projects goals. We did not consider vision focused and computer programming models.

    Two models – medllama2:7b and meditron:7b – were eliminated due to causing repeated infinite loop system locks and providing output that did not follow the prompt instructions. At the time of testing, Medllama did not output any information in JSON format, instead responding with conversational prompts such as:

    \begin{lstlisting}
    ==========================
    Starting query using model medllama2:7b please wait...
    Thank you for providing the patient's detailed information. How would you interpret these findings?

    Time to completion
    Time: -2.584711 seconds
    \end{lstlisting}

    On the other hand, Meditron:7b failed to generate consistent responses. It sometimes used the provided JSON structure and other times hallucinated a new structure or engaged in a conversational manner.

    Gpt-oss:20b was found to not respond when Ollama's structured output setting was enabled. This setting brings a significant boost to the consistency of all model outputs. As such gpt-oss is considered not suitable to the task at hand if it cannot function under the structured output constraint.

    \subsection{Benchmark Dataset Description}

    For testing purposes, the five fictional doctors' notes found in Appendix A are synthesized into a prompt alongside examples if examining few-shot performance, and context vectors in RAG tests. For example, the zero-shot prompt was as follows, with the doctor's note (found in Appendix A) being tested inserted in the section labeled "Doctors note":

    \begin{lstlisting}
    Directions: Provide a response using the provided JSON structure, do not deviate from it, do not create new fields, any time that there is an array multiple datapoints can be added to the array. Use the following JSON structure and if provided context use it for diagnostic code lookup, to process and output the information contained in the doctors note. Include the full original doctors note verbatum in the section labeled original_document. Desired output Json structure: {
        "medical_record": {
            "original_document": "",
            "diagnostic_code": [],
            "diagnosis": [],
        }
    }Doctors note:
    \end{lstlisting}

    The few-shot prompt included the same instructions along with doctors notes 2 and 4 and the related outputs formatted in the following manner with the relevant information placed in the double brackets:

    \begin{lstlisting}
    When provided with this doctors note:
    <<Copy of doctors note found in Appendix A inserted here>>
    The ouput should be:
    {
        "original_document": "<<Copy of doctor's note here>>
    ",
        "diagnostic_codes": ["G51.39"],
        "diagnoses": ["Facial spasm"]
    }
    \end{lstlisting}

    RAG based prompts were based on the following initial prompt, with the context retrieved by LangChain FAISS and Retrieval.

    \begin{lstlisting}
    Context information from relevant documents:
    {context}

    Question: Directions: Provide a response using the provided JSON structure, do not deviate from it, do not create new fields, any time that there is an array multiple datapoints can be added to the array. Use the following JSON structure and if provided context use it for diagnostic code lookup, to process and output the information contained in the doctors note. Include the full original doctors note verbatum in the section labeled original_document.

    Please answer the question based on the context provided above.
    \end{lstlisting}

    LLM responses are saved to a plaintext file with time to response completion is recorded. We then examine in this file the adherence to accuracy of the output JSON structure, time to completion, and correctness of transcription of the doctor's note to the proper field. Extraction of diagnoses and application of diagnostic codes are examined and recorded manually to detect any strange or persistent errors that might cause issues in a medical setting.

    An example of LLM response:

    \begin{lstlisting}
    Starting query using model deepseek-r1:8b please wait...
    {
    "original_document": "79-year-old female presents into the emergency department after a slip and fall that occurred last night while she was trying to get ready for bed. Patient states that she has a history of peripheral neuropathy and she simply tripped on her foot and fell down to hit her face on the dresser. She states she was unable to get up from the ground and get into bed or a couch and spend the night on the ground. Denies any loss of consciousness. She states that she did not allow our housemate to call 911 last night but this morning she was still unable to get up so they called today. Only pain at this time she reports to be in her low back. Denies any loss of consciousness. Any history of anticoagulation or prior head injury.",
    "diagnostic_codes": [
        "I",
        "E",
        "A"
    ],
    "diagnosis": [
        "fall from standing",
        "facial contusion",
        "decomposition of tissue (traumatic rhabdomyolysis)",
        "acute kidney injury"
    ]
    }

    Time to completion
    Time: 7.477762 seconds
    \end{lstlisting}

    In the above response, deepseek-r1:8b is struggling with the task. In this test case (the doctor's note \#3 in Appendix A) deepseek-r1:8b generated diagnostic codes that are nonsensical and diagnoses that only partially match those supplied in the medical note. The correct diagnoses and correct codes would be: 

    \begin{lstlisting}
    "diagnostic_codes": [
        "W01.190A",
        "S00.83XA",
        "E86.0",
        "T79.6XXA",
        "N17.9",
    ],
    "diagnosis": [
        "fall from standing",
        "facial contusion",
        "dehydration",
        "traumatic rhabdomyolysis",
        "acute kidney injury"
    ]
    \end{lstlisting}

    \subsection{Iterative Refinements}
    \label{sec:iterativeRefinement}

    Multiple significant improvements were made in the performance of all models through iterative prompt refinement. In early test runs a field was labeled as

    \begin{lstlisting}
    "chief_complaint_code":""
    \end{lstlisting}

    when it should be labeled

    \begin{lstlisting}
    "diagnostic_codes": []
    \end{lstlisting}

\begin{figure}[htbp]
        \centering
        \includegraphics[width=\linewidth]{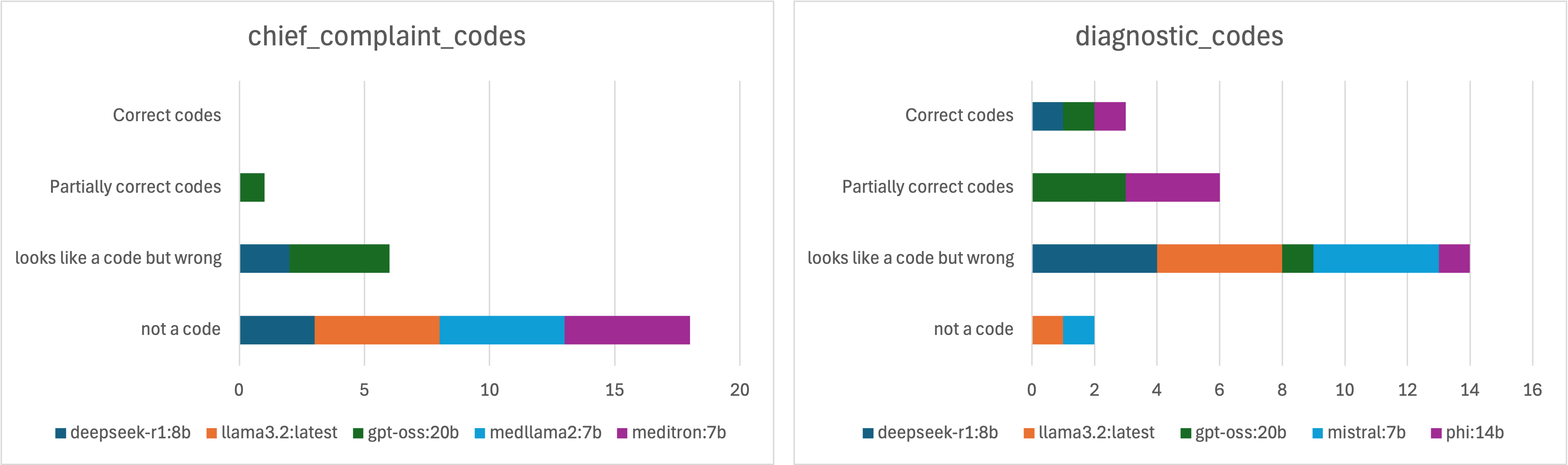}
        \caption{Comparison of performance before and after iterative refinement of JSON terms}
        \label{fig:ChartChiefVDiagnosticCode}
    \end{figure}

    Figure~\ref{fig:ChartChiefVDiagnosticCode} shows the effect of this change in the JSON prompt labels. It led to a marked increase in the presentation of ICD-10 like code responses from the models. With the original label, only gpt-oss:20b returned any results with ICD codes related to the chief complaint while other models left the space blank or inserted random numbers. After changing the label to "diagnostic\_codes"  all models would respond with a codelike structure, with only 2 instances of blank or nonsensical responses.

\begin{figure}[htbp]
        \centering
        \includegraphics[width=\linewidth]{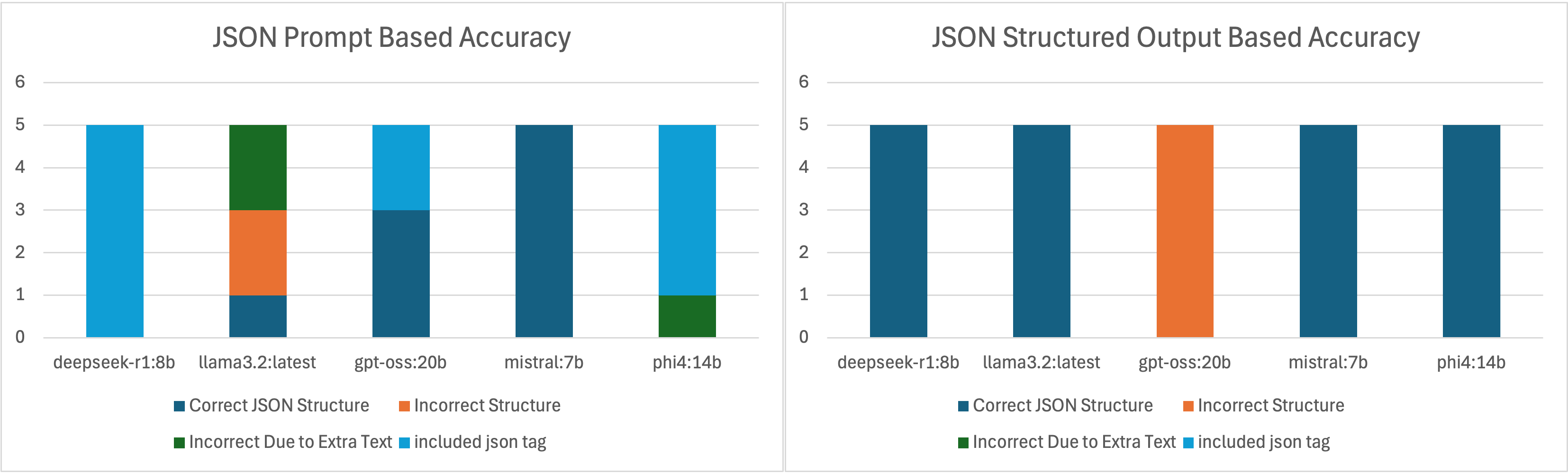}
        \caption{Accuracy of LLM output JSON structure before and after use of Ollama structured output system.}
        \label{fig:jsonPromptVStructureAcc}
    \end{figure}

    The inclusion of Ollama's structured output constraint resulted in a dramatic improvement in output structure compliance. This  inclusion was made after  we assessed the performance of the zero-shot baseline. As Figure~\ref{fig:jsonPromptVStructureAcc} shows, we eliminated all anomalous model responses with flawed JSON based on the structured output system. Unfortunately, it appears that some models are incompatible with the structured output system, in this test gpt-oss consistently failed to produce any output with the structured output system enabled. The performance improvement from the structured output constraint prompts us to disregard gpt-oss from consideration due to this incompatibility.

\begin{figure}[htbp]
        \centering
        \includegraphics[width=\linewidth]{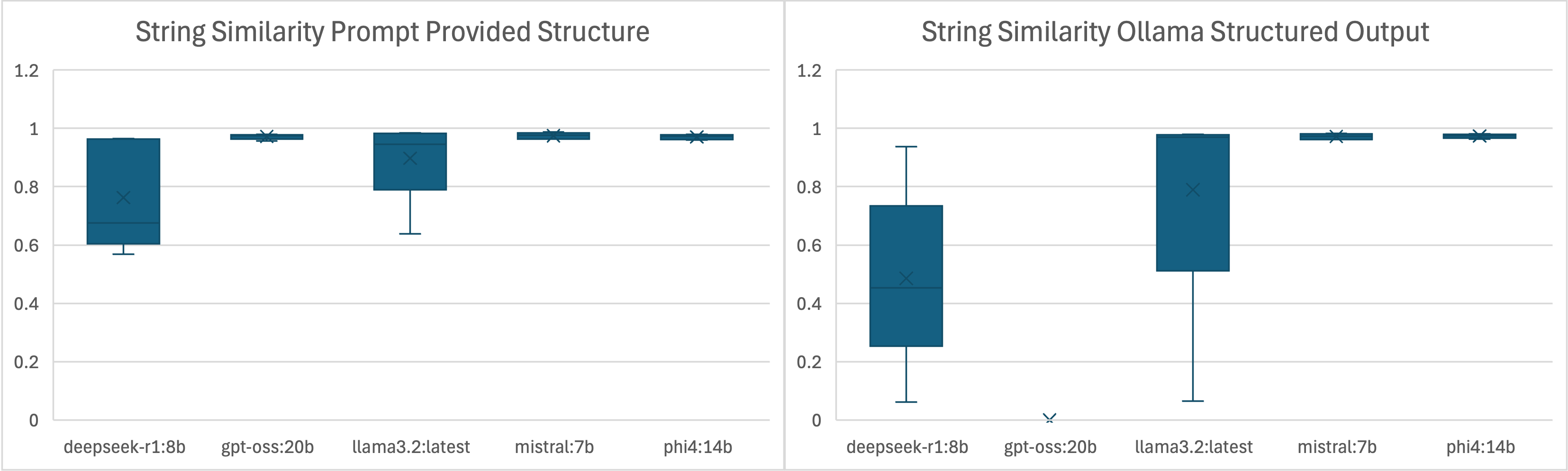}
        \caption{Average string similarity of medical note transcription before and after the inclusion of Ollama structured output system.}
        \label{fig:AvgStrSimPromptvStruct}
    \end{figure}

    Applying this structured constraint system appears to degrade the ability of both deepseek-r1:8b and llama3.2:latest to transcribe the entire medical note into the proper field as shown in Figure~\ref{fig:AvgStrSimPromptvStruct}. Mistral and phi both continue to perform well at this task. 

    Using the structured output system reduced the average runtime of the Deepseek model from approximately 25 seconds to 8 seconds. This may be due to the model no longer streaming the thinking process to output. We did not see any noticeable changes in the runtime of the other models.

    \subsection{Baseline (Zero-shot) Performance}

    Zero-shot prompting is considered our baseline. We evaluate five LLMs without any examples on our task, extracting medical information of the five synthetic doctor's notes and translating it into billing codes. The LLMs were prompted with the doctor's notes inserted into the angled brackets as below:

    \begin{lstlisting}
    Directions: Provide a response using the provided JSON structure, do not deviate from it, do not create new fields, any time that there is an array multiple datapoints can be added to the array. Use the following JSON structure and if provided context use it for diagnostic code lookup, to process and output the information contained in the doctors note. Include the full original doctors note verbatum in the section labeled original_document. Desired output Json structure: {
        "medical_record": {
            "original_document": "",
            "diagnostic_code": [],
            "diagnosis": [],
        }
    }Doctors note:<Doctor's note inserted here>
    \end{lstlisting}

    This prompt was refined over several tests to determine more productive wording for testing.

    \subsubsection{Accuracy of Diagnostic Code Prediction}

\begin{figure}[htbp]
        \centering
        \includegraphics[width=0.8\linewidth]{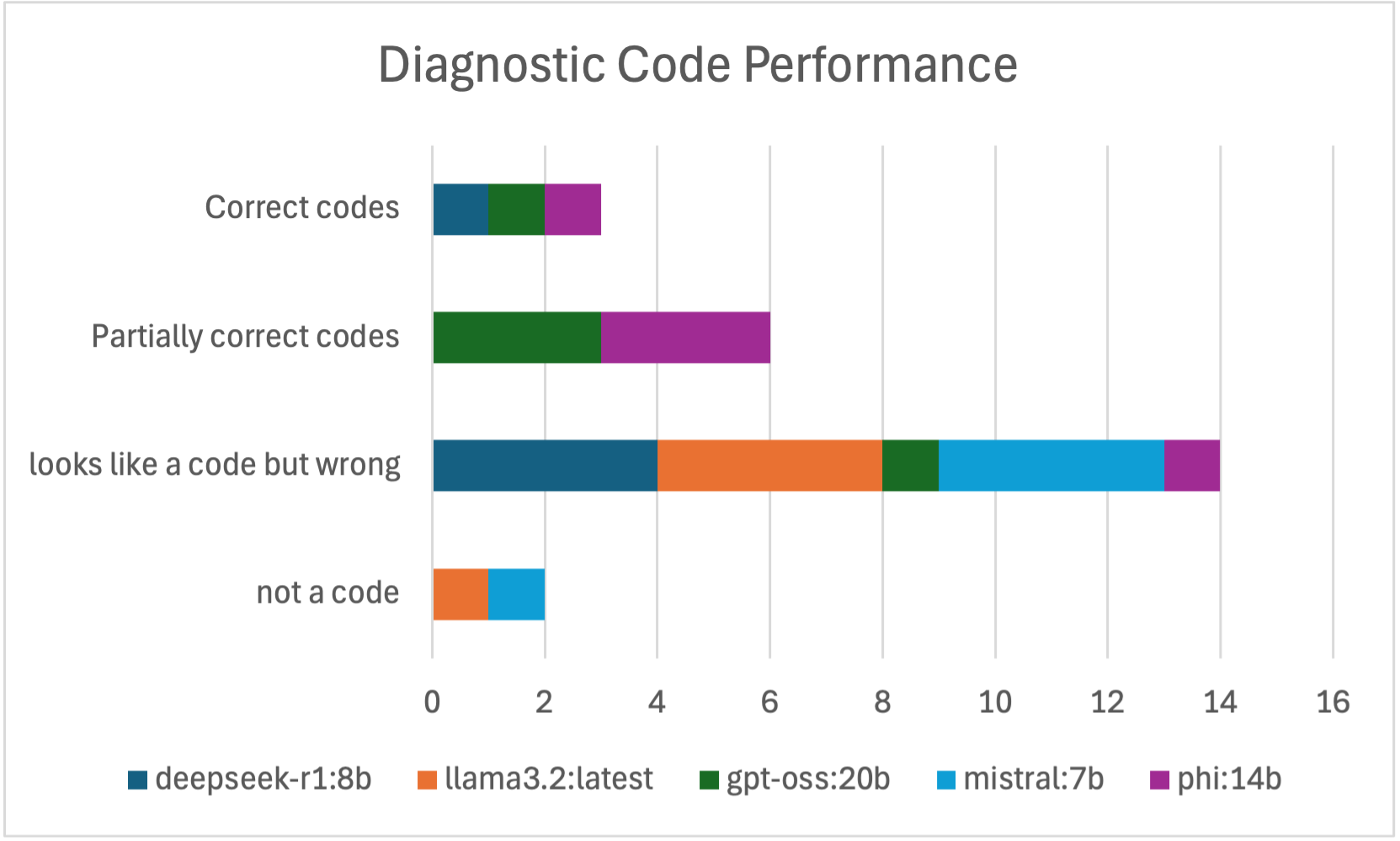}
        \caption{Model successful identification of diagnostic codes in zero-shot prompting experiment. Partially correct codes had at least one code in a prompt correct, but not all. Looks like a code indicates that a code was provided in the proper format, but it does not correlate with any actual ICD-10 code, e.g. Z99.999W}
        \label{fig:diagCodePerfZERO}
    \end{figure}

    Model performance regarding diagnostic codes is suboptimal shown in Figure~\ref{fig:diagCodePerfZERO}. The only codes correctly identified were Streptococcal Pharyngitis: J02.0 identified three times, Dehydration: E86.0 identified twice alongside other incorrect codes, and Acute kidney injury: N17.9 which was identified once properly, and once partially. This results in an unsatisfactory success rate of 3/25 or 12\%. It is also worth noting that the models would consistently add more into the diagnostic\_code field than simply the code, attempting to sometimes add in python style comments, or extra notes that would be problematic for an end case use.

    \subsubsection{Output Structure Compliance}

\begin{figure}[htbp]
        \centering
        \includegraphics[width=0.8\linewidth]{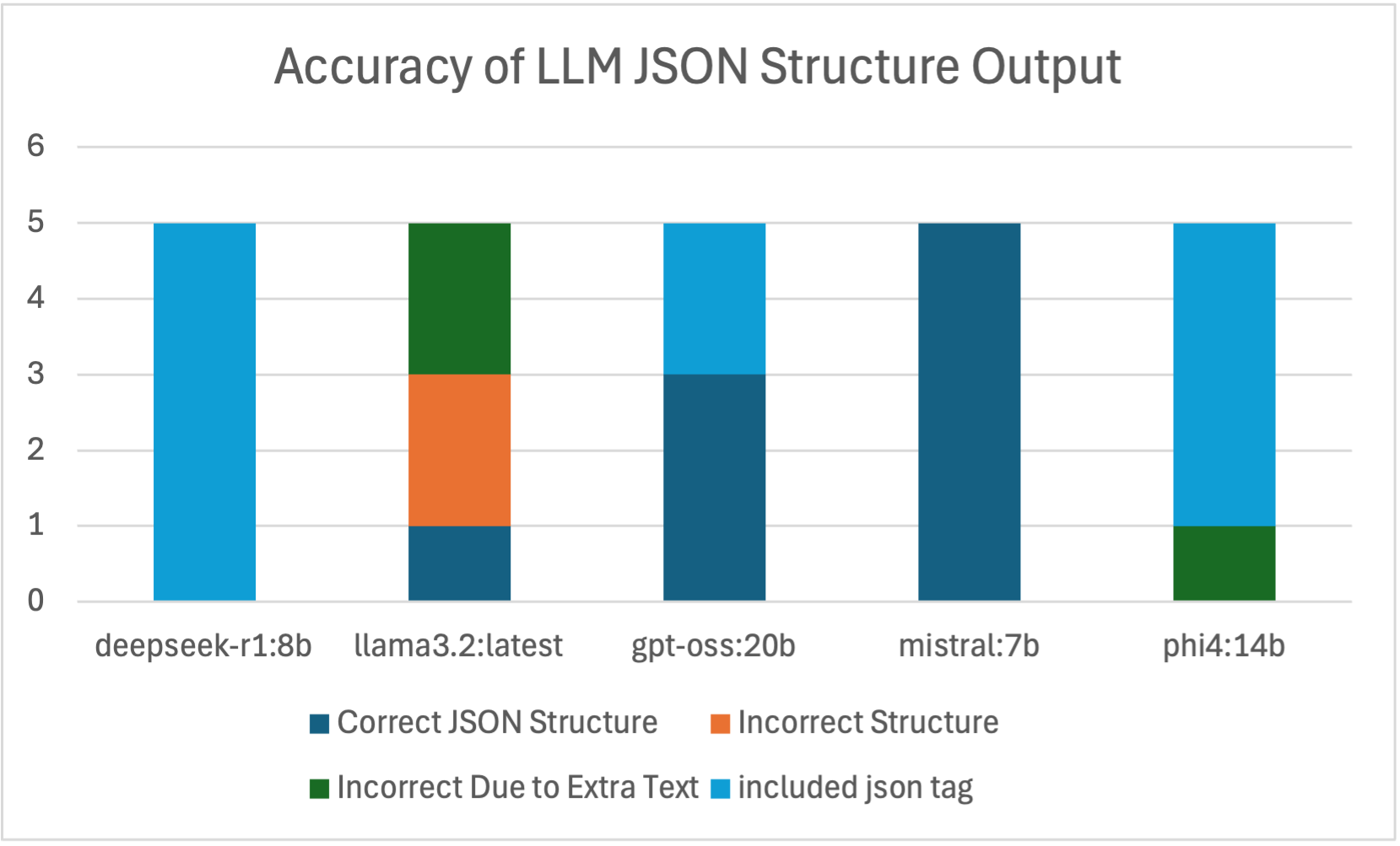}
        \caption{Accuracy of LLM compliance in use of supplied JSON structure.}
        \label{fig:jsonAccuracy}
    \end{figure}

    JSON is a data-only format that does not provide any language for comments \cite{Bray2017}, as such it is critical that the LLM correctly responds with only the JSON structure as rigidly defined in the prompt. Model performance on this metric is shown in Figure~\ref{fig:jsonAccuracy}. A very common error was including a formatting tag around the JSON output of

    \begin{lstlisting}
    '''json
    '''
    \end{lstlisting}

    This was initially considered as a possible feature of the models, however gpt-oss:20b added this formatting tag only occasionally. Other models included additional "helpful" text to the output that invalidated the response. Only llama3.2:latest returned incorrect structures, missing several commas in two of the responses. Mistral performed the most accurately in this test, returning the proper format each time with no extra text, whereas deepseek and phi consistently included the JSON formatting tag.

\begin{figure}[htbp]
        \centering
        \includegraphics[width=0.8\linewidth]{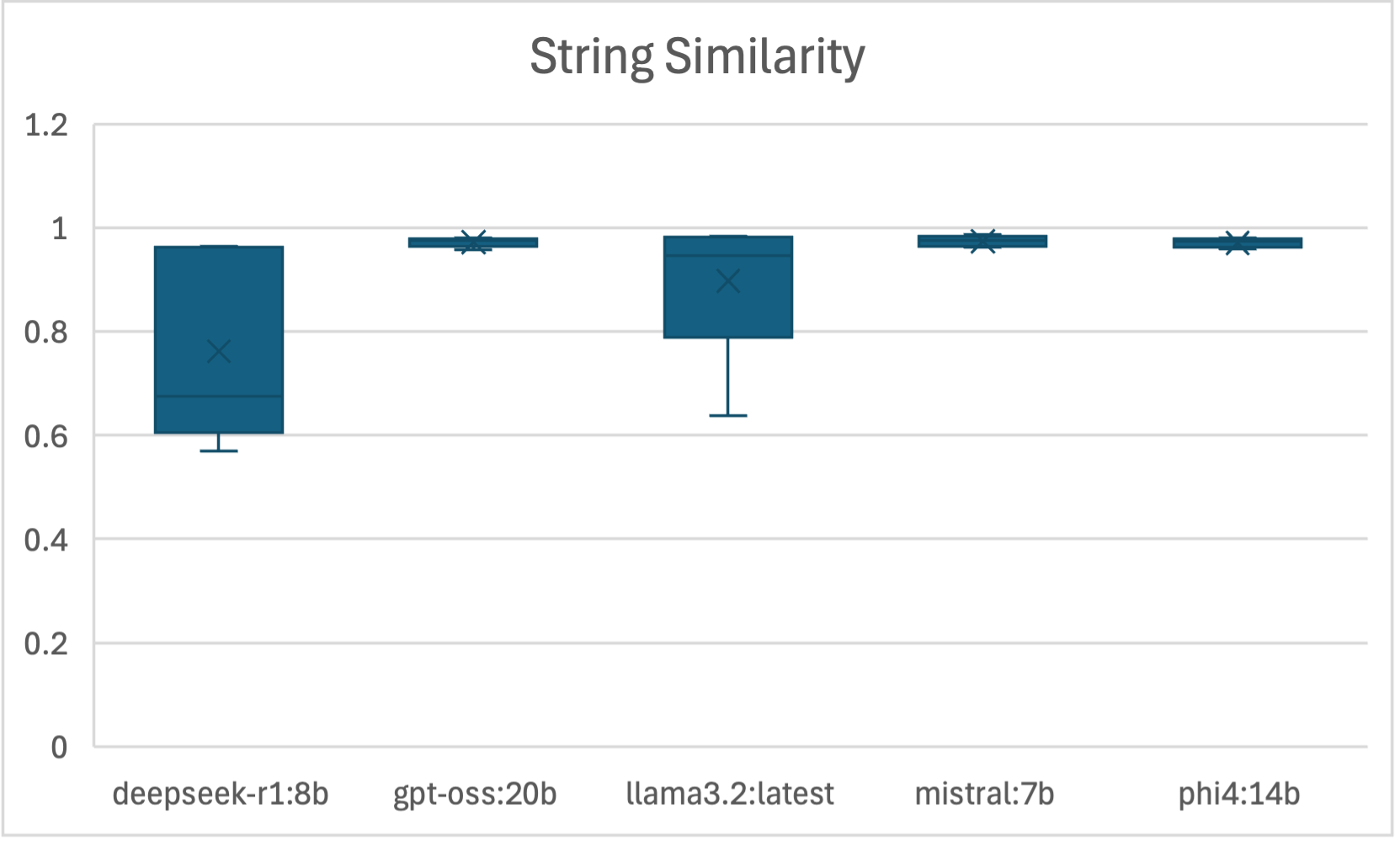}
        \caption{String similarity between prompt supplied doctor's note and LLM transcription into proper JSON field.}
        \label{fig:StringSimZero}
    \end{figure}

    To evaluate more general model compliance, a command was included in the prompt to exactly copy the contents of the doctor's note input into a JSON field. To assess the success of this task the LLM outputs were compared using Pythons native difflib library and determining a simple error rate with a score of 1 being a perfect transcription. After modifying the prompts and dropping non-compliant models as outlined in Section~\ref{sec:iterativeRefinement} the average performance was calculated and compared as shown in Figure~\ref{fig:StringSimZero}. These values could be further refined by running more tests and using more models, but it proves a good metric for determining model compliance on a simple task. In a production level product, it would be best to place the doctor's note in this field directly through scripting instead of using an LLM as it would be faster and more consistent.

    \subsubsection{Response Efficiency}

\begin{figure}[htbp]
        \centering
        \includegraphics[width=0.8\linewidth]{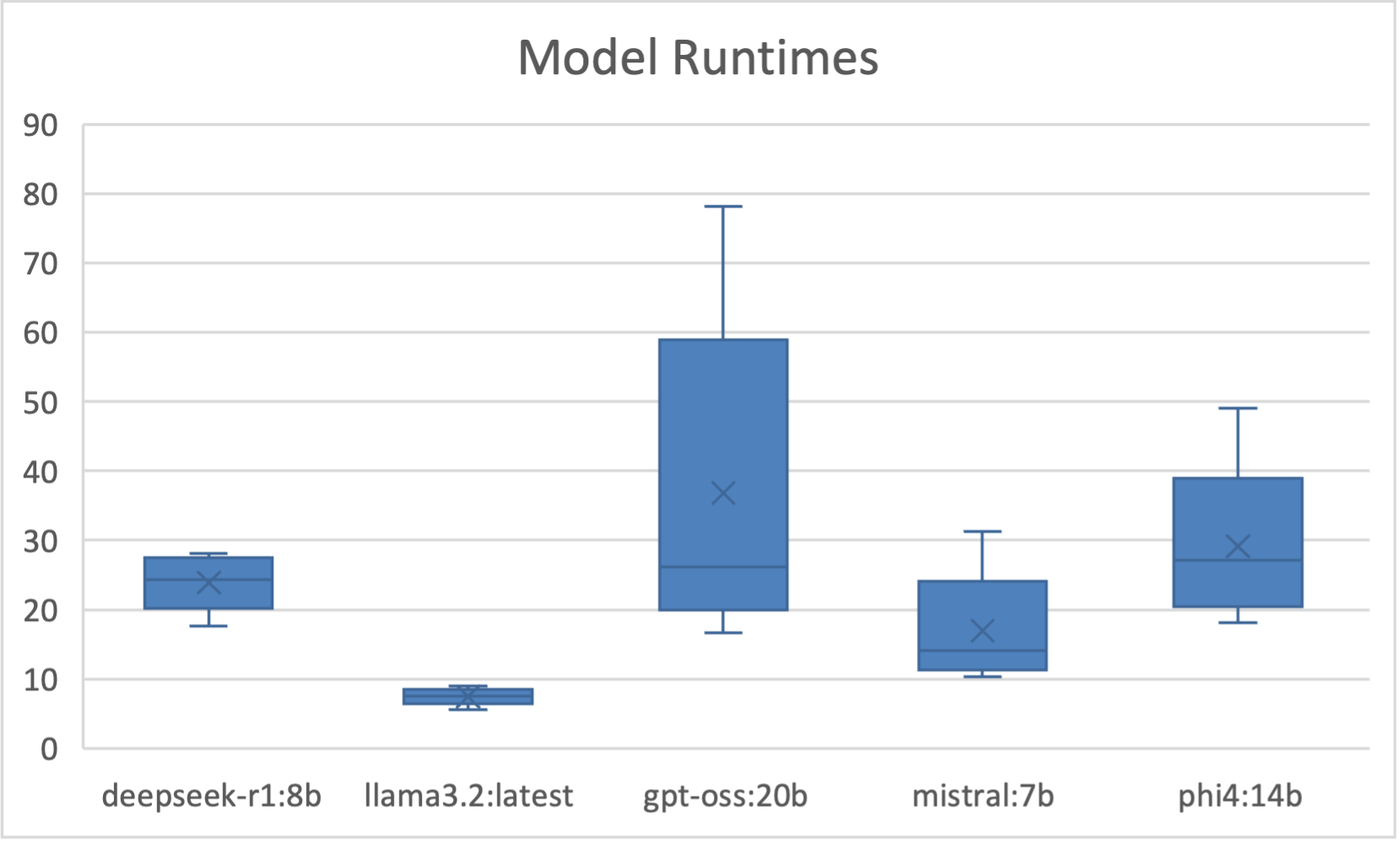}
        \caption{LLM runtimes for zero-shot prompt experiments.}
        \label{fig:modelRuntimesZero}
    \end{figure}

    Model time to complete or runtimes for the zero-shot prompt models were recorded and are summarized in Figure~\ref{fig:modelRuntimesZero}. It is worth noting that llama3.2 was consistently the fastest performer whereas gpt-oss often took the longest time to run with the greatest variance. Mistral – which performed well in the other metrics – was consistently on the faster side and appears to be an ideal candidate.

    \subsection{Few-Shot Prompting Performance}

    To provide annotated examples of outputs to the LLMs in few-shot prompting doctor's notes 2 and 4 were explicitly laid out alongside the exact answers expected from the models. This information was then fed into the LLMs with the instructional prompt and target doctor's note. The JSON structure for these experiments was provided via Ollama's structured output system.

    Example prompt:

    \begin{lstlisting}
    Directions: Provide a response using the provided JSON structure, do not deviate from it, do not create new fields, any time that there is an array multiple datapoints can be added to the array. Use the following JSON structure and if provided context use it for diagnostic code lookup, to process and output the information contained in the doctors note. Include the full original doctors note verbatum in the section labeled original_document. Few shot prompt: When provided with this doctors note:
    <<Full doctors note 4 inserted here>>

    The ouput should be:
    {
        "original_document": "<<Exact copy of the provided doctors note 4 here>",
        "diagnostic_codes": ["G51.39"],
        "diagnoses": ["Facial spasm"]
    }
    \end{lstlisting}

    In testing the doctor's notes that were provided as part of the few-shot prompt were still run as trivial test cases. These trivial cases should return perfect results, if they do not, it may reveal potential issues with parsing.

    The inclusion of the Ollama structured output system brought JSON structure compliance rate to 100\% across all models. However, gpt-oss continued to fail to respond and hence was dropped from consideration. It was replaced at this point with the light weight gemma3:270m model.

    Average string similarity – shown in Figure~\ref{fig:strSimFew} – for the original\_document field continues to perform near 96\% for llama3.2, mistral, and phi4 models. This slight deviation from a 100\% success appears to be due to variance in white space formatting between runs. Deepseek and gemma3.2 performed quite poorly with only 49.5\% and 31.9\% string similarity across tests. We noticed Deepseek consistently summarized the doctor's note instead of transcribing it, printing only "..." for one of the outputs. Gemma3.2, on the other hand, would dramatically summarize the original document such as simplifying the output to only:
    
    \begin{figure}[htbp]
        \centering
        \includegraphics[width=0.8\linewidth]{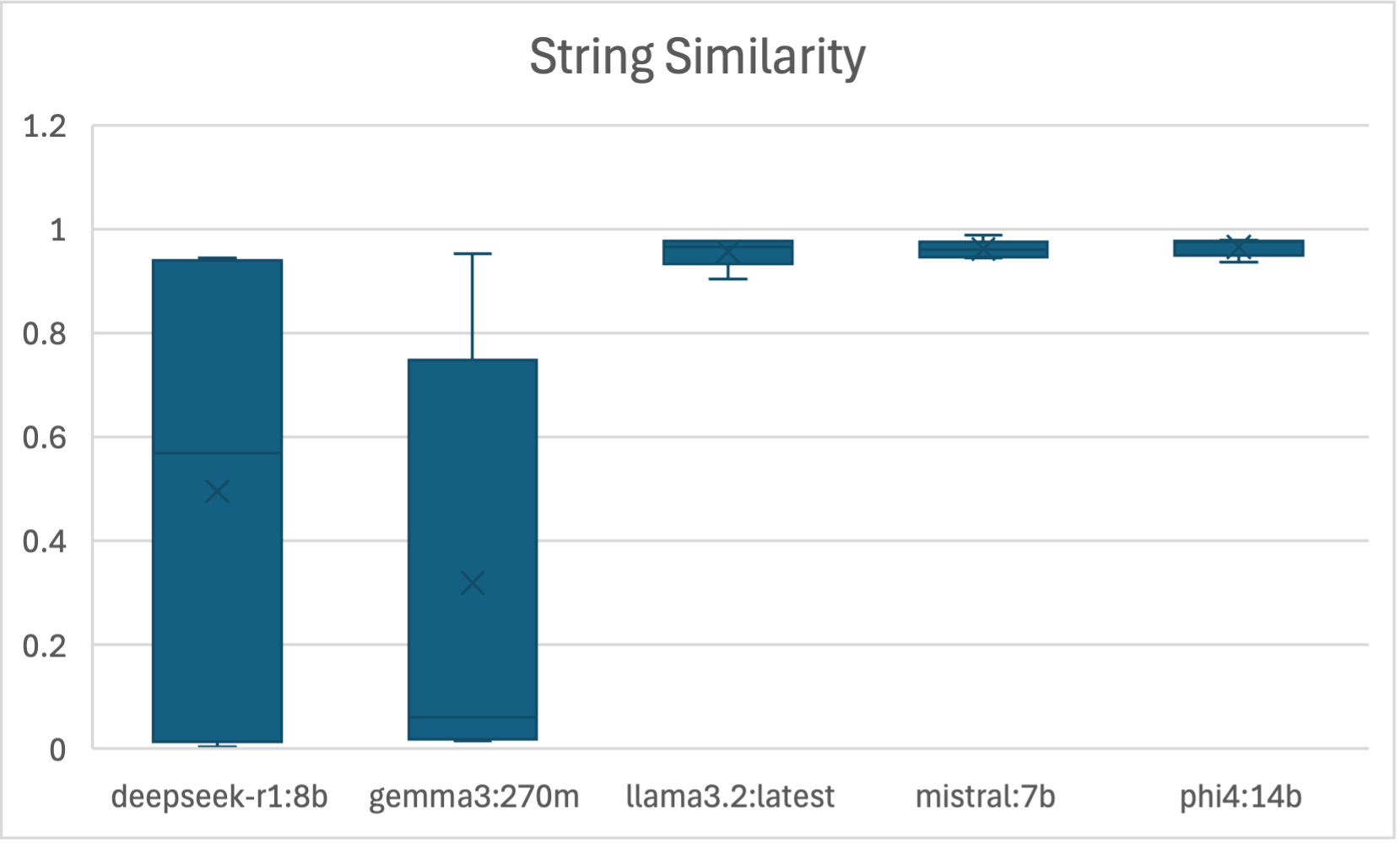}
        \caption{Average string similarity of note transcription by LLM in few-shot scenarios.}
        \label{fig:strSimFew}
    \end{figure}

\begin{figure}[htbp]
        \centering
        \includegraphics[width=0.8\linewidth]{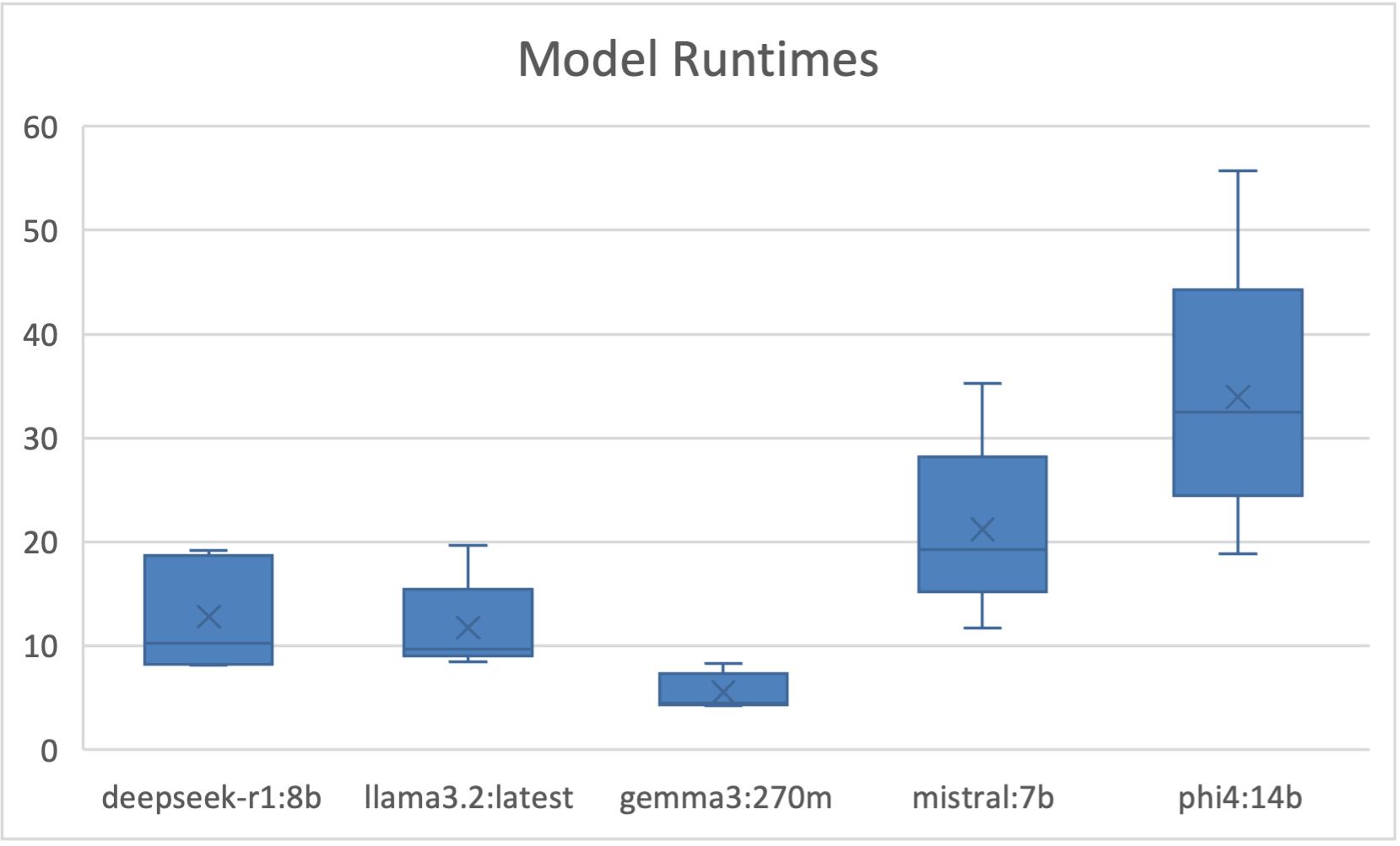}
        \caption{Average model runtimes for few-shot outputs.}
        \label{fig:runtimesFew}
    \end{figure}

    Runtimes remained similar for most models llama3.2 took slightly longer, and the new gemma3.2 model ran very quickly, most likely due to its comparably small size shown in Figure~\ref{fig:runtimesFew}.

    In evaluating the accuracy of output diagnoses and diagnostic codes for the few-shot prompting, it is interesting to note that there are instances where instead of analyzing the provided doctors note, the LLM would simply provide the output of one of the few-shot examples. In our experiment, llama3.2 and gemma3 both exhibited this behavior. 

\begin{figure}[htbp]
        \centering
        \includegraphics[width=0.8\linewidth]{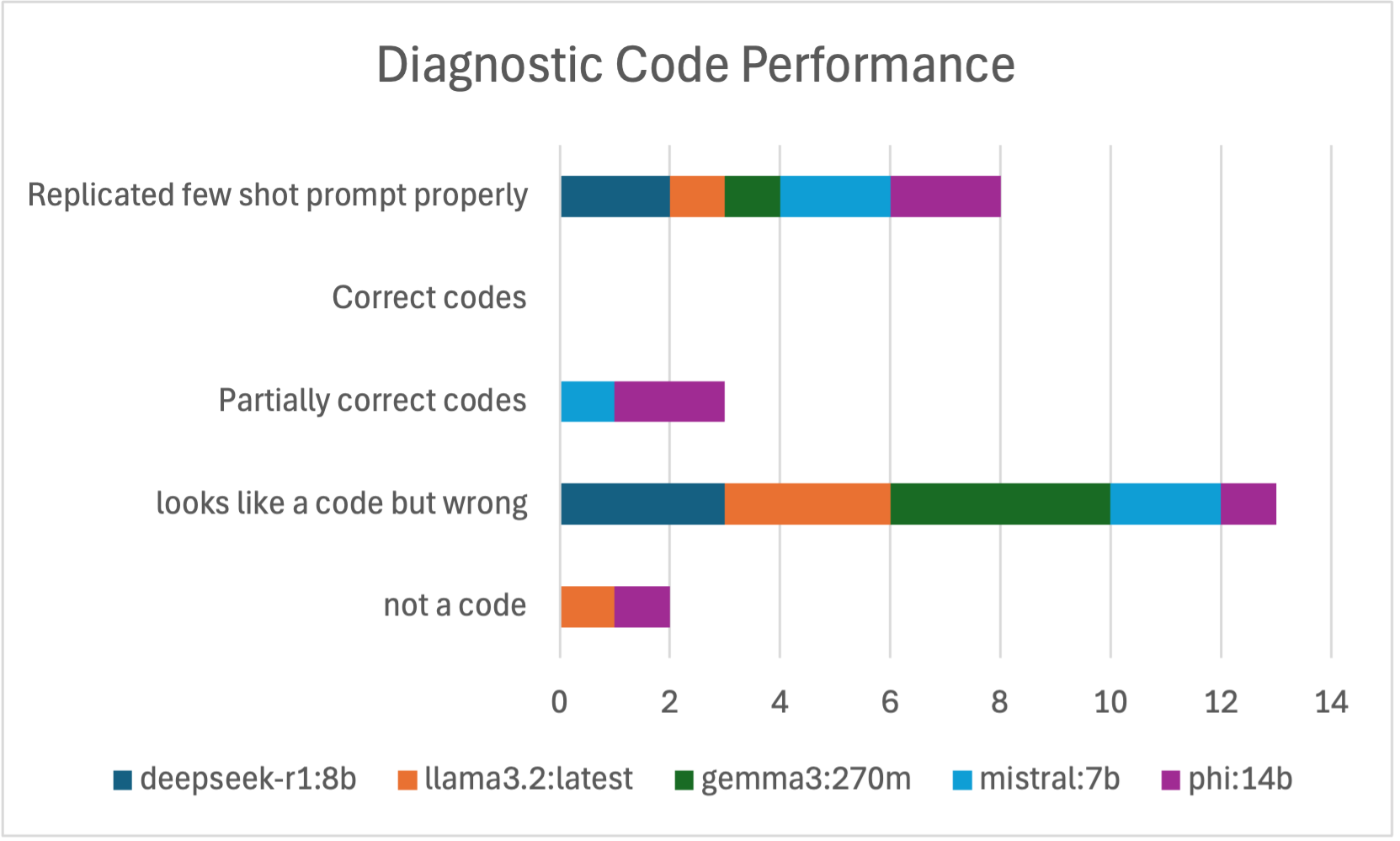}
        \caption{Performance of LLMs when provided with few-shot prompts.}
        \label{fig:diagCodeFew}
    \end{figure}

    Figure~\ref{fig:diagCodeFew} shows the diagnostic code performance accuracy when using the few-shot prompt. Model phi4 almost got the code for streptococcal pharyngitis correct, but it included in the field an attempt at adding a comment to the JSON, nullifying the result. It is interesting to note that models in zero shot prompting tests properly identified the code for streptococcal pharyngitis as J02.0 failed to do so when provided with the few-shot prompt. Mistral and phi4 both got the correct code for acute kidney injury and provided very close values for several other diagnoses.

    Surprisingly, when one compares the results of the few-shot prompting to zero-shot prompting, there is no improvement in the accuracy of the diagnostic codes. In fact, we see a slight decrease in accuracy due to confusion with the few-shot prompt. This lack of difference may be due to already introducing a rigid structure to the LLM outputs using Ollama's structured output system.

\begin{figure}[htbp]
        \centering
        \includegraphics[width=0.8\linewidth]{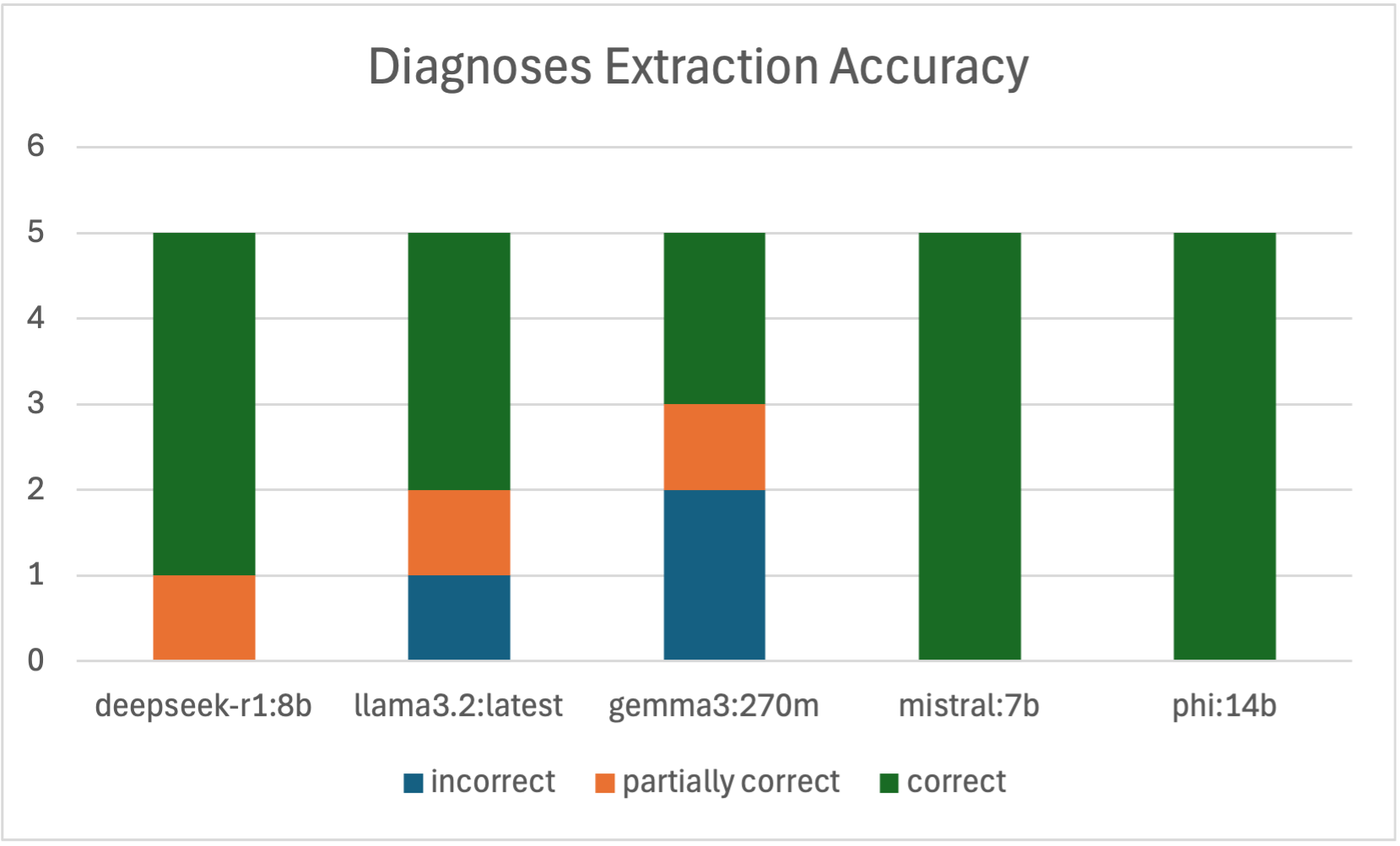}
        \caption{Diagnoses extraction accuracy under few-shot conditions.}
        \label{fig:diagExtractFew}
    \end{figure}

    In terms of the model's ability to extract the diagnoses from the doctor's notes under few-shot conditions, mistral and phi4 continue to perform exceptionally at this task while the other three do present errors as seen in Figure~\ref{fig:diagExtractFew}. 

    \subsection{RAG-Enhanced Prompting Results}

    Similarly to few-shot, LLMs models were prompted but with RAG-enhanced task-relevant examples. We observed Gemma3:270m had not been unable to process the input from the RAG-enhanced prompt structure, repeatedly crashing and causing infinite loops, the very small size of the model is likely to blame so the model was changed to the gemma3:4b version which could process the increased prompt size and context.

    Example Prompt:

    \begin{lstlisting}
    Directions: Provide a response using the provided JSON structure, do not deviate from it, do not create new fields, any time that there is an array multiple datapoints can be added to the array. Use the following JSON structure and if provided context use it for diagnostic code lookup, to process and output the information contained in the doctors note. Include the full original doctors note verbatum in the section labeled original_document. RAG Context Doctors note: <Variable>
    \end{lstlisting}

\begin{figure}[htbp]
        \centering
        \includegraphics[width=0.8\linewidth]{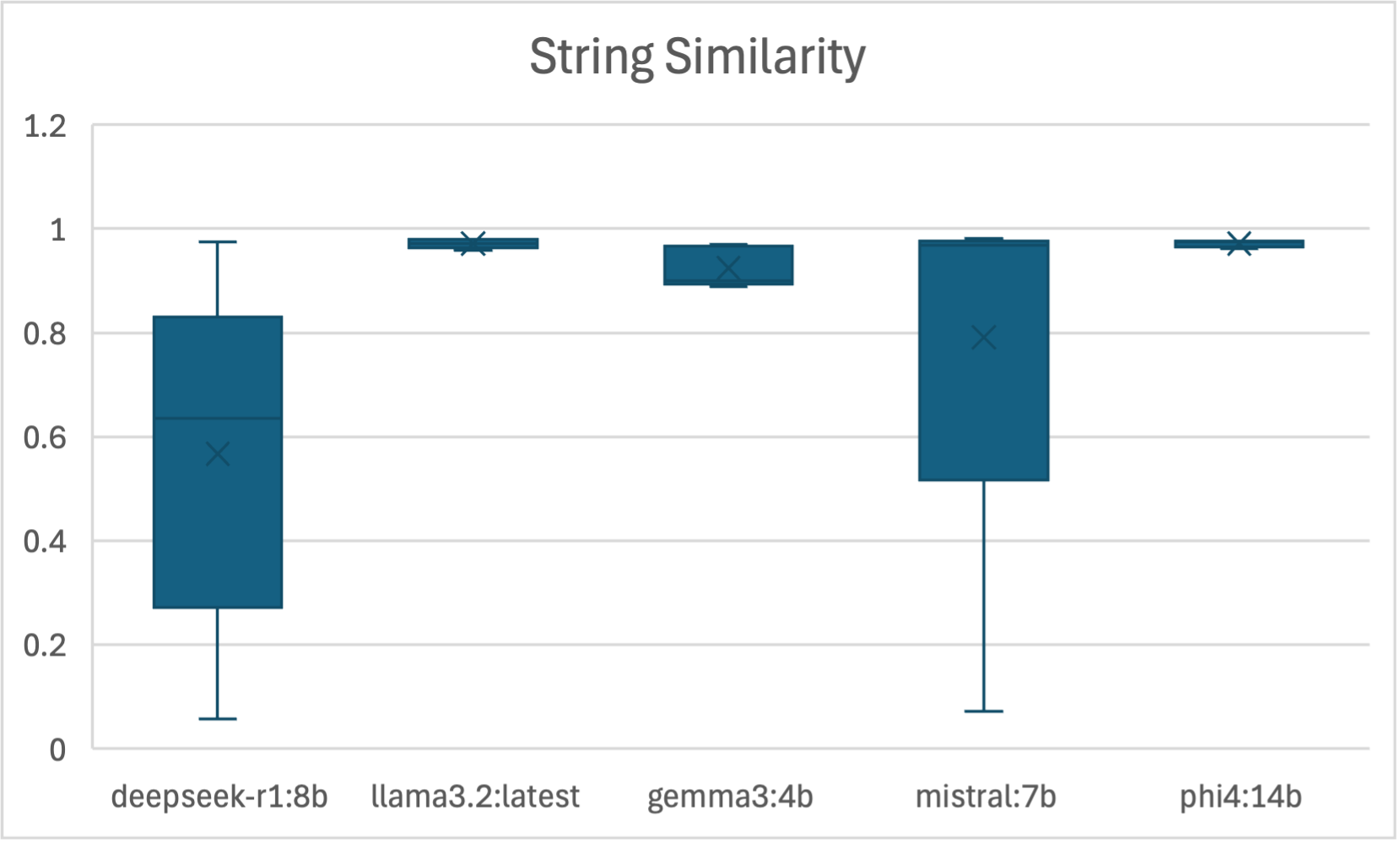}
        \caption{Average string similarity between doctor's note and LLM transcription with RAG-enhanced prompting.}
        \label{fig:strSimRAG}
    \end{figure}

    The transcription of Doctor's notes into the proper JSON field  continued to be consistently high for llama3.2 and phi4 (see Figure~\ref{fig:strSimRAG}). mistral, however, did run into an error and mishandle the task. Deepseek's performance degraded across all metrics when adding the RAG context prompt.

\begin{figure}[htbp]
        \centering
        \includegraphics[width=0.8\linewidth]{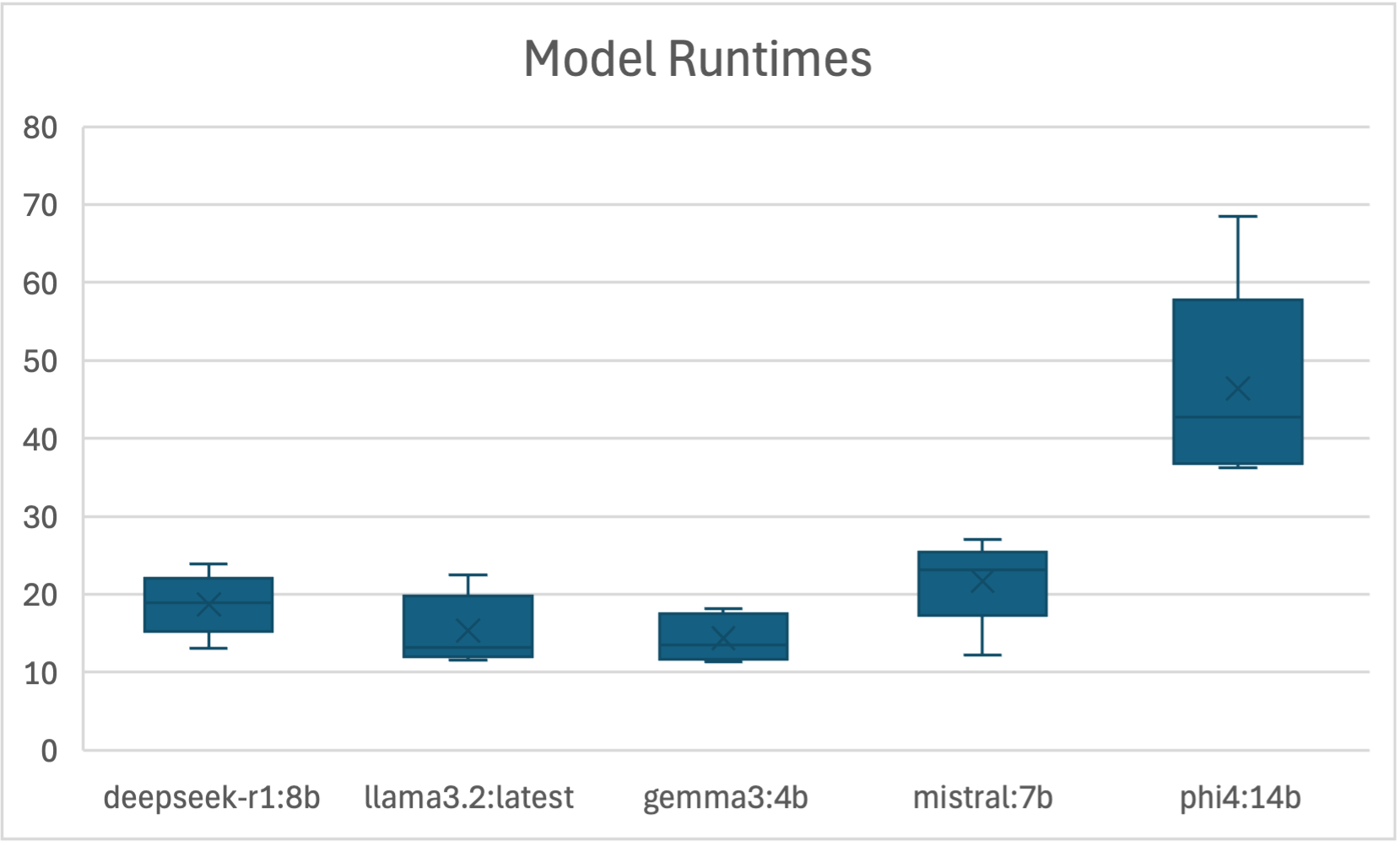}
        \caption{Average model runtime performance with RAG-enhanced prompting.}
        \label{fig:runtimeRAG}
    \end{figure}

    Figure~\ref{fig:runtimeRAG} shows model runtimes are all slightly increased with the addition of the RAG context prompts.

\begin{figure}[htbp]
        \centering
        \includegraphics[width=0.8\linewidth]{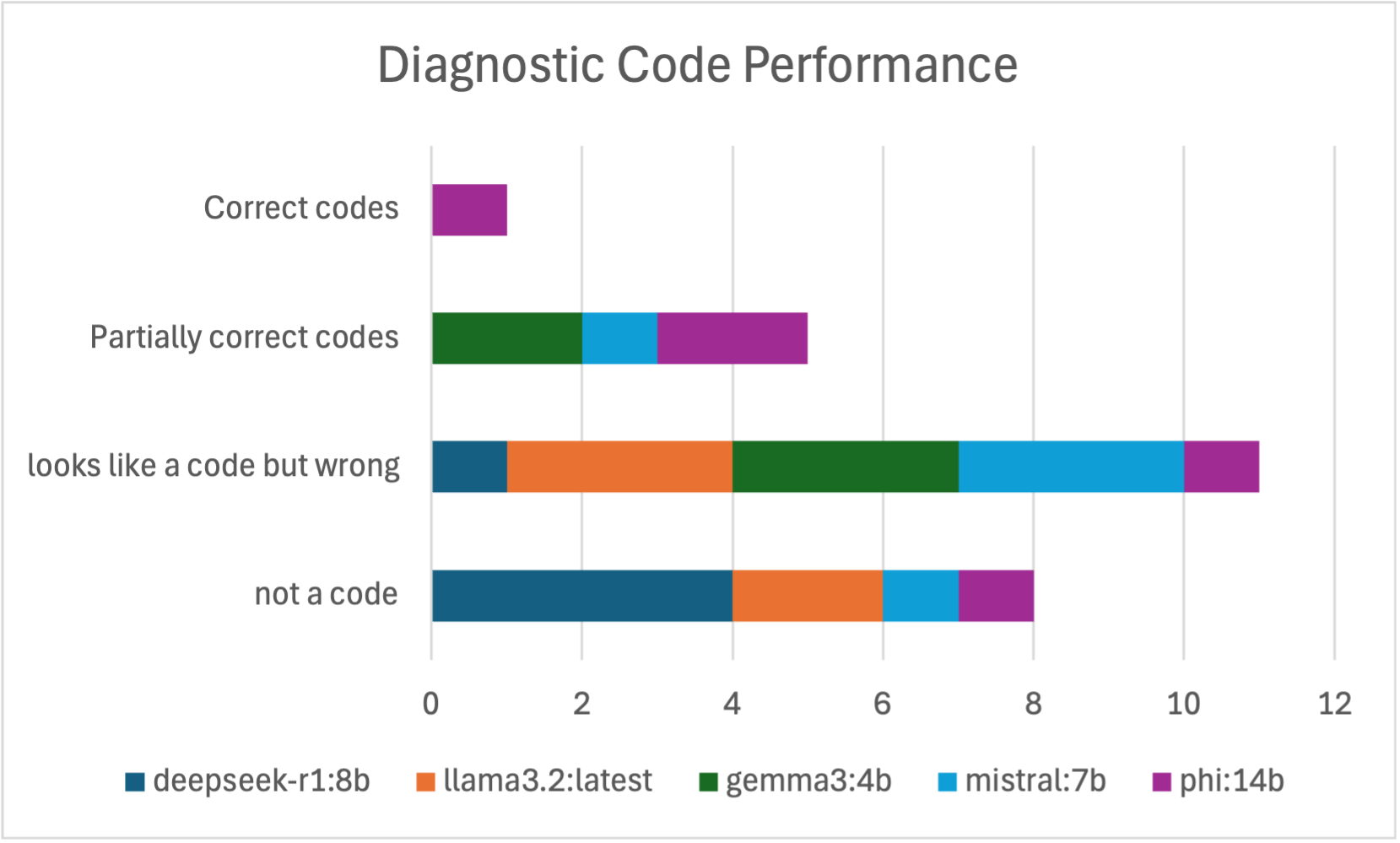}
        \caption{LLM diagnostic code performance using RAG-enhanced prompting.}
        \label{fig:diagPerfRAG}
    \end{figure}

    There were several interesting changes in the model's ability to identify diagnostic codes with the addition of the RAG system. As shown in Figure~\ref{fig:diagPerfRAG}, Gemma identified the code for dyspnea for the first time and phi was very close. Phi produced the correct code for acute low back pain. Llama attempted to create new diagnoses unspecified by the doctor (a severe error). Mistral was partially correct on acute kidney injury, only adding on an improper 7th position code. Phi partially identified the code for facial contusion for the first time. Identifying new codes that the models previously did not recognize is an important improvement, indicating the potential of using the RAG-enhanced prompting  in identifying diagnostic codes to accompany diagnoses. Conversely, Deepseek performed much worse at diagnostic code generation with the introduction of RAG, often leaving the field blank or writing things that were not at all ICD codes.

\begin{figure}[htbp]
        \centering
        \includegraphics[width=0.8\linewidth]{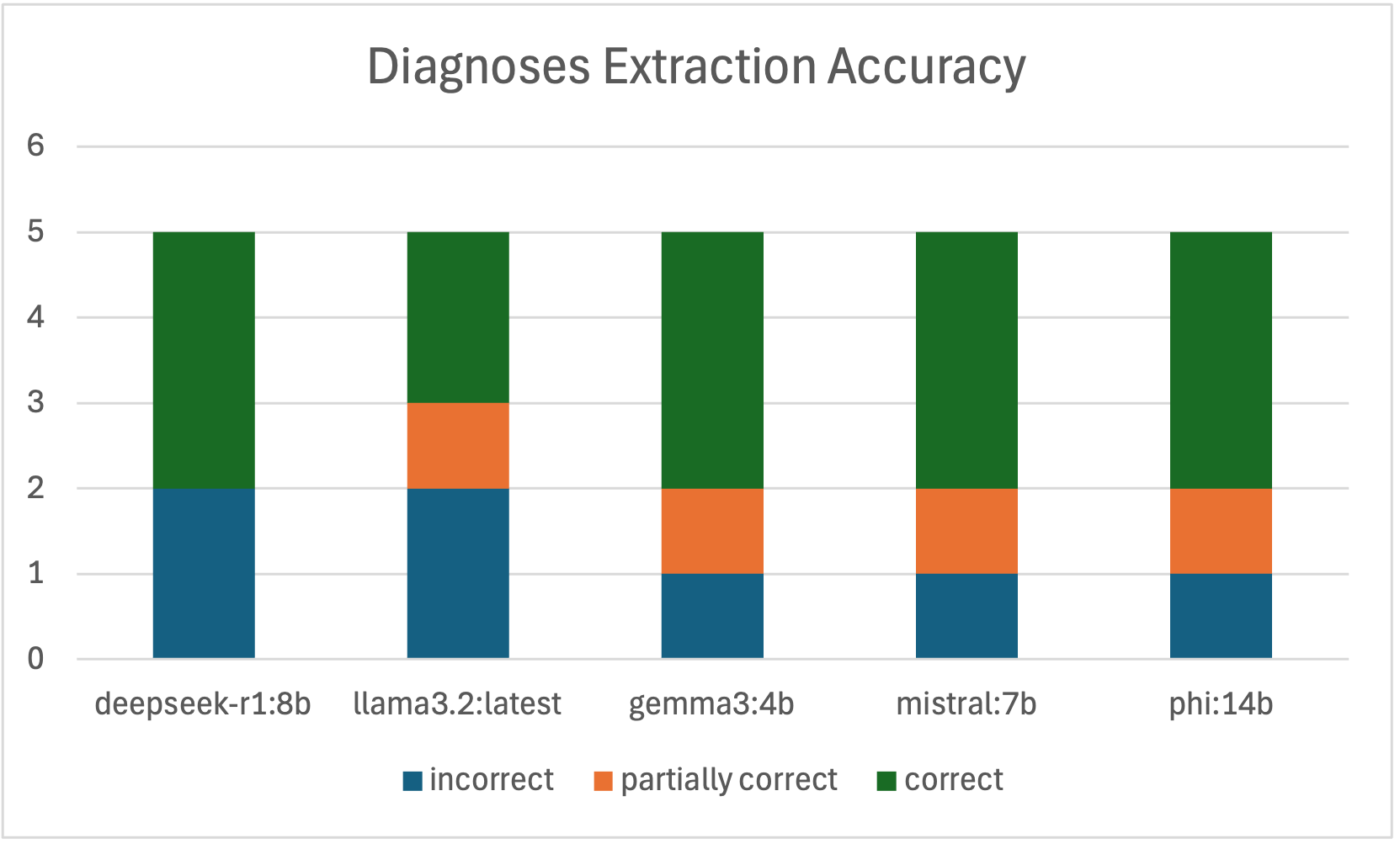}
        \caption{LLM accuracy in extracting diagnoses from doctor's notes with RAG-enhanced prompting.}
        \label{fig:diagExtractRAG}
    \end{figure}

    The introduction of the RAG system decreased diagnoses extraction accuracy (see Figure~\ref{fig:diagExtractRAG}). It may be due to an overabundance of information in the prompt complicating things compared with the simpler prompts when mistral and phi4 performed consistently well. This does indicate that it might be ideal to separate tasks out into several different steps with limited prompts that do one thing at a time as suggested in the multi-agent concept proposed in the methodology section.

    \subsection{Error Analysis}

    Across all experiments several recurring categories of model error were identified. The errors varied in severity and frequency but fell into five groups: (1) structured output errors, (2) transcription errors, (3) diagnostic code hallucinations, (4) prompt misinterpretation and few-shot overfitting, and (5) system-level failures.

    \subsubsection{Structured output errors}

    Despite explicit instructions and providing a rigid JSON schema, many models produced malformed JSON or appended notes and conversational text. A common early error was the insertion of formatting tags such as '''json around outputs. Llama occasionally omitted commas, invalidating responses. The errors may arise from ingrained formatting patterns in model training datasets. Mitigation through Ollama's structured output system was highly effective at correcting these errors.

    \subsubsection{Transcription Errors}

    Including the model compliance test for models to reproduce the doctor's note verbatim within a JSON field proved to be more difficult for LLMs than expected. Models such as deepseek and gemma3:270m frequently summarized and edited the text. These errors may arise from a conflict between the instructions or from a loss of context window. Larger or possibly more instruction-tuned models such as llama3.2, mistral, and phi4  demonstrated consistent adherence, suggesting that model capacity and instruction tuning greatly influence capability. Future mitigations include programmatic insertion of verbatim fields, multistep processing, or possibly multi-agent processing.

    \subsubsection{Diagnostic Code Hallucinations}

    The generation of ICD-10 diagnostic codes remains the most error-prone task, and one of the most important goals of the research. In zero-shot tests, models included fabricated or partial codes, appended commentary that is not compatible with JSON, or inserted non-ICD-like values such as indexes. Even when the RAG system was introduced, some models continued to produce corrupt entries and invent diagnoses. These failures may be attributed to limited internal representations of ICD-10 structure and the lack of domain specific training. The RAG pipeline improved the identification of several previously unseen codes, indicating that exposure to reference material improves performance; however, this effect was model-dependent and unreliable for some LLMs. Mitigation strategies include deeper examination of the RAG context processing, splitting diagnosis identification and code lookup into an ordered process, and enforcing stricter output validation.

    \subsubsection{Prompt Misinterpretation and Few-Shot Overfitting}

    Unexpectedly, few-shot prompting decreased performance on several tasks. Multiple times llama3.2 and gemma3 ignored the target doctor's note and instead reproduced answers from the few-shot prompt. This indicates a pattern-matching bias or overfitting to the provided context. Strict separators, clearer instructions, or modifying the style of prompt formatting may reduce these failures.

    \subsubsection{System-Level Failures}

    Medllama2:7b, meditron:7b and gemma3:270m frequently entered infinite loops, refused to follow instructions and produced conversational replies instead of structured output. Gemma3:270m was completely incapable of handling larger prompt sizes, especially when including the RAG-enhanced tests. These failures appear related to model architecture and scale limitations. Mitigation includes selecting larger models, and continued evaluation of other models for performance stability.

    \subsection{Summary of Findings}

    \begin{itemize}
        \item \textbf{Structured output enforcement is highly effective} — Ollama's
        structured output system eliminated malformed JSON responses for all
        compliant models and occasionally reduced runtime, marking it valuable
        for deterministic data extraction.
    
        \item \textbf{Prompt engineering and terminology significantly impact
        performance} — minor changes such as pluralizing fields or changing a field
        name from \texttt{chief\_complaint\_code} to \texttt{diagnostic\_code}
        produced major improvements in model behavior.
    
        \item \textbf{Few-shot prompting was unexpectedly ineffective} — instead of
        improving accuracy, few-shot examples induced pattern-copying errors and
        degraded diagnostic code performance.
    
        \item \textbf{Gpt-oss is incompatible with Ollama structured output} and
        cannot be used reliably.
    
        \item \textbf{Deepseek exhibits major performance degradation} in RAG
        prompting contexts.
    
        \item \textbf{Model size constrains task capabilities} — \texttt{Gemma3:270M}
        was unable to process larger RAG-based prompts, while \texttt{Gemma3:4B}
        handled them successfully.
    
        \item \textbf{RAG-enhanced prompting improves some diagnostic code
        identification and merits further examination} — however, it decreases
        diagnosis extraction accuracy, suggesting the need for multi-step or
        agent-based workflows.
    \end{itemize}

    \section{Discussion}

    The results of this research highlights some of the complexities at play when dealing with data privacy, large language model (LLM) architecture, and the problems inherent to the administrative medical system. The proposed system successfully demonstrated a privacy-preserving pipeline for clinical data, however limited successful code generation and response errors reveal important insights into attempts to use open-weight LLMs.

    One of the most important constraints for this work is the requirement of on-device network isolated processing to protect Controlled Unclassified Information (CUI) and Patient Health Information (PHI). The prioritization of local execution and limited access to large server sized hardware limited model selection to relatively small models in the range of 1 billion to 20 billion parameters. This resulted in a noticeable gap in performance capabilities in 'reasoning' and in context window size.

    Our experiments show that smaller models excel at speed and simple tasks with short context windows, such as identifying and transcribing diagnoses terms. However, they struggle with semantic reasoning tasks that would map diagnoses (e.g., "Sore throat") to specific ICD-10-CM codes (e.g. "J02.0"). Large commercial agentic systems may be able to handle this task due to network access and trillion-parameter architectures, however the privacy constraints and desire for an accessible open-source system preclude that option. While privacy-preserving AI is currently feasible for data structuring it requires further development and human-in-the-loop verification for clinical use.

    One of the most unexpected findings of this study was the degradation of performance under few-shot prompting conditions. Standard prompt engineering literature suggests that providing examples via in-context learning typically improves model outputs. However, in our tests the inclusion of full JSON example texts caused model regression. This may be due to attention drive from small context windows common in smaller models. When presented with token-heavy JSON examples in prompts, the models appeared to overfit the content of the examples to the output rather than the logic of the task. Instead of analyzing the new note, models occasionally hallucinated information by copying data points from the few-shot examples into the final output. This finding suggests that for medical coding tasks on local devices, explicit schema enforcement (such as Ollama's structured output system) is a more effective control mechanism that traditional few-shot prompting.

    The integration of Retrieval-Augmented Generation (RAG) produced conflicting results. While it enabled the identification of previously unseen diagnostic codes, it frequently confused models and created errors in the data outputs. This implies a noise to signal issue within the context window. Injecting large chunks of raw ICD-10CM definitions into the prompt may have saturated the model's attention mechanisms. The 7B parameter models struggled to distinguish between the instructions, the doctor's note, and the RAG context. This prompt approach proved too demanding for the hardware constraints. This supports the argument for a multistep or multi-agent architecture, where the tasks of clinical extraction and code lookup are separated into distinct workflows.

    It is important to acknowledge that the failures of this system are not solely technological; they are also a reflection of the complexity of medical billing. As Don Norman argues in The Design of Everyday Things, if a user (in this case, an LLM) struggles with a design, the fault lies with the designer, not user error. The ICD coding system is not well-designed for logical inference; it is an administrative taxonomy filled with arbitrary codes, rules, and specificities (e.g., specific codes for "struck by a duck" and "struck by a turkey") that defy intuition. We are asking a probabilistic model to solve a deterministic bureaucratic problem. The difficulty the LLMs faced in generating accurate codes reflects the unnecessary cognitive burden being placed on human physicians every day. If the underlying coding system utilized a more logical, human-centered design it is likely that automated systems would perform with significantly higher accuracy.

    \section{Conclusion}

    This thesis set out to explore the viability of a privacy-focused, local LLM system for automating the extraction and coding of doctor's notes into medical information. Through the development of a RAG-based pipeline and the evaluation of multiple open-source models, this work offers a clear assessment of current capabilities and limitations of on-device medical AI and paints a potential path forward.

    The primary contribution of this work is the validation of a secure offline pipeline using Ollama, LangChain, and containerized environments. We demonstrated that enforcing JSON schemas via "structured outputs" is a highly effective method for making local LLMs usable in clinical workflows, achieving near 100\% format compliance across compliant models. Furthermore, in creating a small benchmark of synthetic medical notes, the research addresses the critical scarcity of open, privacy-safe test data, providing a foundation for future research. The benchmark dataset serves as a proof-of-concept but is insufficient for statistical significance and requires expansion to include other fields of practice and medical writing styles.

    Hardware constraints restricted testing to consumer-grade silicon, preventing the evaluation of larger (100B+) models that might possess the reasoning capabilities required for higher coding accuracy. Future work could include more powerful hardware and advancements in AI science and hardware technology will open more paths. 

    Future work should pivot away from large prompts that contain the whole problem and move towards multi-agent or multistep architectures. A next viable step is to split the pipeline into a "clerk" that will extract and sort factual data from the medical note, and a "coder" that will focus only on looking up the appropriate codes in the provided RAG context. It is also possible that replacing RAG with low-rank adaptation (LoRA) fine-tuning could help embed the language of the ICD codes into the model weights reducing context window noise. Finally, to move beyond academic theory, future iterations can focus on integrating with medical standards such as FHIR to integrate with electronic health record systems such as Epic. A clean and user-friendly UI is also a critical component of any software for use by the public.

    While this research concludes that fully automated, unsupervised medical coding via local LLMs is not reliable enough in this state, the concept of Assisted Coding is viable and there is potential for development of a product through continued testing and iterative improvements. The system proves effective at the drudgery of data formatting and extraction. By shifting the physician's role away from "data entry clerk" and "ICD code lookup agent" towards "verifier" – then further back to "physician" – we can significantly reduce the administrative burdens that drive provider burnout. These findings align with the broader conclusion of this thesis that, although fully autonomous coding currently remains out of reach for smaller open-weight models, a human-in-the-loop framework leveraging zero-shot, few-shot, or RAG-based prompting strategies represents the most practical near-term solution. The work contributes an open-source, reproducible local LLM architecture and benchmark dataset designed specifically for extracting and translating medical information into ICD-10-CM codes, supporting continued innovation in privacy-preserving clinical AI.

    The technology is there; the challenge lies in refining the implementation and optimizing prompting strategies to reduce hallucinations, mitigate context-window saturation, and improve diagnostic-code accuracy in local models.

\bibliography{references}

@article{Babaiha2024,
  title    = {Rationalism in the face of GPT hypes: Benchmarking the output of large language models against human expert-curated biomedical knowledge graphs},
  volume   = {5},
  issn     = {26673185},
  doi      = {10.1016/j.ailsci.2024.100095},
  journal  = {Artificial Intelligence in the Life Sciences},
  author   = {Babaiha, Negin Sadat and Rao, Sathvik Guru and Klein, Jürgen and Schultz, Bruce and Jacobs, Marc and Hofmann-Apitius, Martin},
  year     = {2024},
  month    = june,
  pages    = {100095},
  language = {en}
}

@book{Bray2017,
  title       = {The JavaScript Object Notation (JSON) Data Interchange Format},
  url         = {https://www.rfc-editor.org/info/rfc8259},
  doi         = {10.17487/RFC8259},
  number      = {RFC8259},
  institution = {RFC Editor},
  author      = {Bray, T.},
  year        = {2017},
  month       = dec,
  pages       = {RFC8259},
  language    = {en}
}

@article{Cascella2024,
  title    = {The Breakthrough of Large Language Models Release for Medical Applications: 1-Year Timeline and Perspectives},
  volume   = {48},
  issn     = {1573-689X},
  doi      = {10.1007/s10916-024-02045-3},
  number   = {1},
  journal  = {Journal of Medical Systems},
  author   = {Cascella, Marco and Semeraro, Federico and Montomoli, Jonathan and Bellini, Valentina and Piazza, Ornella and Bignami, Elena},
  year     = {2024},
  month    = feb,
  pages    = {22},
  language = {en}
}

@article{Chang2024,
  title    = {Use of SNOMED CT in Large Language Models: Scoping Review},
  volume   = {12},
  issn     = {2291-9694},
  doi      = {10.2196/62924},
  journal  = {JMIR Medical Informatics},
  author   = {Chang, Eunsuk and Sung, Sumi},
  year     = {2024},
  month    = oct,
  pages    = {e62924},
  language = {en}
}

@article{Han2025,
  title    = {Enhancing semantical text understanding with fine-tuned large language models: A case study on Quora Question Pair duplicate identification},
  volume   = {20},
  issn     = {1932-6203},
  doi      = {10.1371/journal.pone.0317042},
  number   = {1},
  journal  = {PLOS ONE},
  author   = {Han, Sifei and Shi, Lingyun and Tsui, Fuchiang (Rich)},
  editor   = {Mustafa, Sohaib},
  year     = {2025},
  month    = jan,
  pages    = {e0317042},
  language = {en}
}

@article{Mesko2023,
  title    = {Prompt Engineering as an Important Emerging Skill for Medical Professionals: Tutorial},
  volume   = {25},
  issn     = {1438-8871},
  doi      = {10.2196/50638},
  journal  = {Journal of Medical Internet Research},
  author   = {Meskó, Bertalan},
  year     = {2023},
  month    = oct,
  pages    = {e50638},
  language = {en}
}

@article{Na2018,
  title   = {Feasibility of Reidentifying Individuals in Large National Physical Activity Data Sets From Which Protected Health Information Has Been Removed With Use of Machine Learning},
  volume  = {1},
  issn    = {2574-3805},
  doi     = {10.1001/jamanetworkopen.2018.6040},
  number  = {8},
  journal = {JAMA Network Open},
  author  = {Na, Liangyuan and Yang, Cong and Lo, Chi-Cheng and Zhao, Fangyuan and Fukuoka, Yoshimi and Aswani, Anil},
  year    = {2018},
  month   = dec,
  pages   = {e186040}
}

@misc{Oaklander2016,
  title        = {Why Doctors Are Burned Out by Busywork},
  url          = {https://time.com/4383979/doctor-burnout-electronic-health-records/},
  abstractnote = {Electronic health records and digital clerical work are strongly linked to burnout},
  journal      = {TIME},
  author       = {Oaklander, Mandy},
  year         = {2016},
  month        = june,
  language     = {en}
}

@article{Packhauser2022,
  title     = {Deep learning-based patient re-identification is able to exploit the biometric nature of medical chest X-ray data},
  volume    = {12},
  rights    = {2022 The Author(s)},
  issn      = {2045-2322},
  doi       = {10.1038/s41598-022-19045-3},
  number    = {1},
  journal   = {Scientific Reports},
  publisher = {Nature Publishing Group},
  author    = {Packhäuser, Kai and Gündel, Sebastian and Münster, Nicolas and Syben, Christopher and Christlein, Vincent and Maier, Andreas},
  year      = {2022},
  month     = sept,
  pages     = {14851},
  language  = {en}
}

@article{Peng2023,
  title    = {A study of generative large language model for medical research and healthcare},
  volume   = {6},
  issn     = {2398-6352},
  doi      = {10.1038/s41746-023-00958-w},
  number   = {1},
  journal  = {npj Digital Medicine},
  author   = {Peng, Cheng and Yang, Xi and Chen, Aokun and Smith, Kaleb E. and PourNejatian, Nima and Costa, Anthony B. and Martin, Cheryl and Flores, Mona G. and Zhang, Ying and Magoc, Tanja and Lipori, Gloria and Mitchell, Duane A. and Ospina, Naykky S. and Ahmed, Mustafa M. and Hogan, William R. and Shenkman, Elizabeth A. and Guo, Yi and Bian, Jiang and Wu, Yonghui},
  year     = {2023},
  month    = nov,
  pages    = {210},
  language = {en}
}

@inbook{Podder2025,
  address    = {Treasure Island (FL)},
  title      = {SOAP Notes},
  rights     = {Copyright © 2025, StatPearls Publishing LLC.},
  callnumber = {NBK482263},
  url        = {http://www.ncbi.nlm.nih.gov/books/NBK482263/},
  booktitle  = {StatPearls},
  publisher  = {StatPearls Publishing},
  author     = {Podder, Vivek and Lew, Valerie and Ghassemzadeh, Sassan},
  year       = {2025},
  language   = {eng}
}

@inbook{Quercia2024,
  title     = {MedFrenchmark, a Small Set for Benchmarking Generative LLMs in Medical French},
  rights    = {https://creativecommons.org/licenses/by-nc/4.0/},
  isbn      = {978-1-64368-533-5},
  url       = {https://ebooks.iospress.nl/doi/10.3233/SHTI240486},
  doi       = {10.3233/SHTI240486},
  booktitle = {Studies in Health Technology and Informatics},
  publisher = {IOS Press},
  author    = {Quercia, Amandine and Zaghir, Jamil and Lovis, Christian and Gaudet-Blavignac, Christophe},
  editor    = {Mantas, John and Hasman, Arie and Demiris, George and Saranto, Kaija and Marschollek, Michael and Arvanitis, Theodoros N. and Ognjanović, Ivana and Benis, Arriel and Gallos, Parisis and Zoulias, Emmanouil and Andrikopoulou, Elisavet},
  year      = {2024},
  month     = aug
}

@article{Rocher2019,
  title     = {Estimating the success of re-identifications in incomplete datasets using generative models},
  volume    = {10},
  rights    = {2019 The Author(s)},
  issn      = {2041-1723},
  doi       = {10.1038/s41467-019-10933-3},
  number    = {1},
  journal   = {Nature Communications},
  publisher = {Nature Publishing Group},
  author    = {Rocher, Luc and Hendrickx, Julien M. and de Montjoye, Yves-Alexandre},
  year      = {2019},
  month     = july,
  pages     = {3069},
  language  = {en}
}

@article{Sinsky2016,
  title     = {Allocation of Physician Time in Ambulatory Practice: A Time and Motion Study in 4 Specialties},
  volume    = {165},
  issn      = {0003-4819},
  doi       = {10.7326/M16-0961},
  number    = {11},
  journal   = {Annals of Internal Medicine},
  publisher = {American College of Physicians},
  author    = {Sinsky, Christine and Colligan, Lacey and Li, Ling and Prgomet, Mirela and Reynolds, Sam and Goeders, Lindsey and Westbrook, Johanna and Tutty, Michael and Blike, George},
  year      = {2016},
  month     = dec,
  pages     = {753–760}
}

@article{Tai2023,
  title    = {Association of physician burnout with perceived EHR work stress and potentially actionable factors},
  volume   = {30},
  rights   = {https://creativecommons.org/licenses/by/4.0/},
  issn     = {1067-5027, 1527-974X},
  doi      = {10.1093/jamia/ocad136},
  number   = {10},
  journal  = {Journal of the American Medical Informatics Association},
  author   = {Tai-Seale, Ming and Baxter, Sally and Millen, Marlene and Cheung, Michael and Zisook, Sidney and Çelebi, Julie and Polston, Gregory and Sun, Bryan and Gross, Erin and Helsten, Teresa and Rosen, Rebecca and Clay, Brian and Sinsky, Christine and Ziedonis, Douglas M and Longhurst, Christopher A and Savides, Thomas J},
  year     = {2023},
  month    = sept,
  pages    = {1665–1672},
  language = {en}
}

@article{Taylor2024,
  title    = {Developing healthcare language model embedding spaces},
  volume   = {158},
  issn     = {09333657},
  doi      = {10.1016/j.artmed.2024.103009},
  journal  = {Artificial Intelligence in Medicine},
  author   = {Taylor, Niall and Schofield, Dan and Kormilitzin, Andrey and Joyce, Dan W. and Nevado-Holgado, Alejo},
  year     = {2024},
  month    = dec,
  pages    = {103009},
  language = {en}
}

@article{Veen2023,
  title     = {Clinical Text Summarization: Adapting Large Language Models Can Outperform Human Experts},
  rights    = {https://creativecommons.org/licenses/by/4.0/},
  doi       = {10.21203/rs.3.rs-3483777/v1},
  publisher = {In Review},
  author    = {Veen, Dave Van and Uden, Cara Van and Blankemeier, Louis and Delbrouck, Jean-Benoit and Aali, Asad and Bluethgen, Christian and Pareek, Anuj and Polacin, Malgorzata and Reis, Eduardo Pontes and Seehofnerova, Anna and Rohatgi, Nidhi and Hosamani, Poonam and Collins, William and Ahuja, Neera and Langlotz, Curtis and Hom, Jason and Gatidis, Sergios and Pauly, John and Chaudhari, Akshay},
  year      = {2023},
  month     = oct
}

@misc{Wilcox2022,
  title    = {How Soon Will the United States Adopt ICD-11? - Find-A-Code Medical Coding and Billing Articles},
  url      = {https://www.findacode.com/articles/how-soon-will-the-united-states-adopt-icd-11-36983.html},
  journal  = {Find-A-Code},
  author   = {Wilcox, Aimee},
  year     = {2022},
  month    = feb,
  language = {en}
}

@article{Yang2023,
  title     = {TCM-GPT: Efficient Pre-training of Large Language Models for Domain Adaptation in Traditional Chinese Medicine},
  rights    = {Creative Commons Attribution 4.0 International},
  url       = {https://arxiv.org/abs/2311.01786},
  doi       = {10.48550/ARXIV.2311.01786},
  publisher = {arXiv},
  author    = {Yang, Guoxing and Shi, Jianyu and Wang, Zan and Liu, Xiaohong and Wang, Guangyu},
  year      = {2023}
}

@article{Zhang2025,
  title    = {Revolutionizing Health Care: The Transformative Impact of Large Language Models in Medicine},
  volume   = {27},
  issn     = {1438-8871},
  doi      = {10.2196/59069},
  journal  = {Journal of Medical Internet Research},
  author   = {Zhang, Kuo and Meng, Xiangbin and Yan, Xiangyu and Ji, Jiaming and Liu, Jingqian and Xu, Hua and Zhang, Heng and Liu, Da and Wang, Jingjia and Wang, Xuliang and Gao, Jun and Wang, Yuan-geng-shuo and Shao, Chunli and Wang, Wenyao and Li, Jiarong and Zheng, Ming-Qi and Yang, Yaodong and Tang, Yi-Da},
  year     = {2025},
  month    = jan,
  pages    = {e59069},
  language = {en}
}

@misc{OllamaBlog2024,
  author       = {Ollama},
  title        = {Structured Outputs},
  howpublished = {\url{https://ollama.com/public/structured-outputs}},
  year         = {2024},
  month        = dec,
  note         = {Accessed: 2025-06-14}
}

@misc{i2MC,
  author       = {AMCICoding},
  title        = {Introduction to Medical Coding (i2MC)},
  howpublished = {\url{https://amcicoding.thinkific.com/products/courses/i2MC}},
  note         = {Accessed: 2025-06-14}
}

@misc{LLM02,
  author       = {OWASP},
  title        = {LLM02: Insecure Output Handling},
  howpublished = {\url{https://genai.owasp.org/llmrisk/llm02-insecure-output-handling/}},
  year         = {2025},
  note         = {OWASP GenAI Top 10; Accessed: 2025-06-14}
}

@misc{cms2025,
  author       = {{Centers for Medicare \& Medicaid Services}},
  title        = {Overview of Coding \& Classification Systems},
  howpublished = {\url{https://www.cms.gov/cms-guide-medical-technology-companies-and-other-interested-parties/coding/overview-coding-classification-systems}},
  note         = {Accessed: 2025-11-1},
  year         = {2025}
}

@misc{Burnout2025,
  title        = {What is physician burnout?},
  url          = {https://www.ama-assn.org/practice-management/physician-health/what-physician-burnout},
  howpublished = {\url{https://www.ama-assn.org/practice-management/physician-health/what-physician-burnout}},
  journal      = {American Medical Association},
  author       = {American Medical Association},
  year         = {2025},
  month        = may,
  language     = {en},
  note         = {Accessed: 2025-6-14}
}
\clearpage

\appendix
\section{Fictional Medical Notes}
\label{app:fictionalMedNotes}
Appendix A reproduces clinical notes verbatim, preserving the original formatting as presented to the model.
\subsection{Patient Note 1 - Pediatric}
\label{app:fictionalNote1}
\begin{lstlisting}
    Subjective
Chief Complaint: Sore throat 
History of Present illness: 15-year-old female patient presents with 2 days of worsening sore throat. She first noticed the sore throat upon waking yesterday morning and says that the pain has gotten worse despite taking over the counter acetaminophen and using cough drops. Today she has had pain when she swallows and says it almost feels like food is "getting stuck." She has been able to eat and drink, although she can only take soft food. She is not sure whether she has had any fevers, but she did feel chilled last night and again this morning. She denies any swollen glands or difficulty moving her neck. 
Past Medical History:  Seasonal allergic rhinitis, worse in the spring. 
Past Surgical History: Appendectomy at age 5, no complications
Family History: Father with type 2 DM, mother with rheumatoid arthritis
Social History: She lives with her mother (parents are divorced) and younger sister. She is not sexually active. She does not smoke, vape, or use alcohol, marijuana or other recreational/illicit drugs. 
LMP:  ~ 1 week ago 
Medications: She takes a daily multivitamin for teens
Allergies: No known allergies to medications or foods. 
Review of Systems:
	General: + for fatigue and malaise, +chills, no fevers; no weight gain or loss. 
HEENT: + for sore throat and difficulty swallowing. No congestion or rhinorrhea. No ear pain. No vision changes. No Neck stiffness. 
	Respiratory: No cough, wheezing, or shortness of breath
	Cardiac: No chest pain, palpitations or syncope
Gastrointestinal: + mild nausea. No vomiting, diarrhea or constipation. No abdominal pain. Normal appetite. 
	Endocrine: No polyuria or polydipsia
	Neurology: +headache. No dizziness, numbness or tingling. 
	Hematology/Oncology:  No swollen glands or easy bruising or bleeding
	Skin:  No rash or other skin changes
	Musculoskeletal:  No joint pain or muscle pain
	Psychiatric:  No major stressors or mood changes

Objective
Vital signs:  HR 102 bpm, RR 14 bpm, BP 108/72, Temp 101.4F, PulseOx 98%, Height, Weight, BMI, BMI%
Physical Examination:
	General: Well nourished, well developed. Tired appearing but in no distress
	Head: Normal appearance, no trauma or lesions
Neck: Supple. + tender anterior lymphadenopathy bilaterally . No thyromegaly or masses
	Eyes: PERRL, no conjunctival erythema or discharge
Ears:  Normal external canals bilaterally, tympanic membranes with normal landmarks & no erythema or bulging
Nose: Mild erythema and edema of the turbinates bilaterally with scant clear rhinorrhea. No lesions.
Mouth/Throat: Dentition grossly intact. No oral lesions. Posterior oropharynx is erythematous with 2+ tonsillar hypertrophy bilaterally and patchy, scattered exudates. + palatal petechiae.
Lungs: Normal work of breathing. Lungs are clear to auscultation bilaterally with no wheezing, rales, rhonchi. Normal air excursion. 
	Heart: Regular rate and rhythm. Mildly tachycardic. No murmurs, rubs, or gallops
Abdomen: Non-distended. Normal bowel sounds in all 4 quadrants. Non-tender to palpation with no organomegaly or masses
	Skin: No rash on exposed skin
	Musculoskeletal: Moving all four extremities normally
	Extremities: Warm and well perfused, with normal pulses and normal capillary refill
	Neuro:  Appropriately alert and interactive for age. Normal gait, balance and speech
	Psych:  Mood and affect are euthymic.

Laboratory Tests: Rapid strep test + 
Radiology Tests: Not indicated 

Assessment: 
Otherwise healthy 15 year old female with acute streptococcal pharyngitis. 

Plan: 
-	1000mg Amoxicillin daily. X 10 days. 
-	Ibuprofen 400-600mg every 6-8 hours as needed for pain (take with food)
-	Increase fluids & use throat lozenges for comfort
-	May return to school after 12-24 hours of antibiotics, as long as she is fever-free off of antipyretics and feeling better
-	Counselled on the importance of completing antibiotics 
-	Return to clinic as needed if symptoms worsen or fail to improve within 24-48 hours

\end{lstlisting}
\subsection{Diagnosis and ICD-10-CM Code for Patient Note 1}
Streptococcal Pharyngitis: J02.0

\subsection{Patient Note 2 - ER}
\label{app:fictionalNote2}
\begin{lstlisting}
    HPI:
    Is an 81 year-old female presenting into the emergency department after having a slip and fall at home. Patient states that she was on her way to the bathroom, she slipped on the bathroom floor and fell and hit her head. She says that she's been having back pain for the past one week and it's worsened at this time. The pain is constant, non-radiating. Pain worsens with any attempted movement. Denies any bladder incontinence. No loss of conscience. Not any anticoagulants. Denies any chest pain or shortness of breath.
    
    Past medical history:
    Denies any past medical history.
    Not on any medication's.
    
    Physical examination:
    Patient awake, alert, well appearing, no acute distress.
    Head is normal is normocephalic, atraumatic
    Normal peripheral perfusion with regular rate rhythm.
     Nonlabored respirations with equal chest rise, clear to auscultation 
    No focal CTL tenderness to palpation
    Focal tenderness to right greater trochanter and left mid tibia. NVI distal extremities. 
    
    medical decision-making:
    81-year-old presenting after a slip and fall.
    Differential diagnosis includes and cranial hemorrhage, contusion, fracture.
    
    Un reassessment patient remains hemodynamically stable and with non-focal neurologic exam. Plane films of the right hip and the left leg are negative for acute fracture, CT scan of the cervical spine, and the lumbar spine are negative for acute fracture or other acute traumatic injury. CT of the head shows a questionable asymmetric thickening of the inferior falx.
    
    Called to discuss with on-call neurosurgeon, in agreement with admission to our facility, repeat had CT at six hours. Patient and family are updated on all results and comfort a plan of care. Called and discussed with hospitalist for admission.
    
    I personally viewed an interpreted patient's EKG, chest, x-ray, plain films, CT imaging, laboratory results.
    
    Disposition - admit for further observation and repeat head CT
    
    Diagnosis
    Fall from standing
    Acute low back pain
    Abnormal imaging result
    
\end{lstlisting}
\subsection{Diagnosis and ICD-10-CM Code for Patient Note 2 - ER}

Fall from standing: Z91.81, More correct and specific: History of falling W01.0XXA

Acute low back pain: M54.50

Abnormal imaging results: R93.0

\subsection{Patient Note 3 - ER}
\label{app:fictionalNote3}
\begin{lstlisting}
    Chief complaint trip and fall

    HPI 79-year-old female presents into the emergency department after a slip and fall that occurred last night while she was trying to get ready for bed. Patient states that she has a history of peripheral neuropathy and she simply tripped on her foot and fell down to hit her face on the dresser. She states she was unable to get up from the ground and get into bed or a couch and spend the night on the ground. Denies any loss of consciousness. She states that she did not allow our housemate to call 911 last night but this morning she was still unable to get up so they called today. Only pain at this time she reports to be in her low back. Denies any loss of consciousness. Any history of anticoagulation or prior head injury.
    
    Past medical history includes hypertension, hyperlipidemia, peripheral neuropathy, vitamin D, deficiency, diabetes.
    
    Physical exam.
    Patient awake alert, oriented in no acute distress.
    Extract clear muscles are intact with no diplopia.
    There is an abrasion above the left orbit and there is ecchymosis surrounding the left orbit and tracking down the left side of the face with associated infra orbital tenderness.
    Tolerate full range of motion in the neck with no midline, cervical thoracic or lumbar tenderness, palpation step off or deformity.
    Normal peripheral perfusion.
    Non-labored, respirations with equal bilateral chest rise
    
    
    Medical decision-making
    79 year-old female, presenting into the emergency department after slip and fall at home last night.
    Differential diagnosis includes intracranial, hemorrhage, fracture, contusion, AI, dehydration, non-exertional rhabdomyolysis
    
    Imaging is significant for no acute traumatic injuries, patient feels improved after IV hydration, her creatinine is slightly elevated, CK is slightly elevated. Patient is reserved to hear in the emergency department for four hours and remains hemodynamically stable. Basic metabolic panel and creatinine kinase is repeated And both are slightly improved after hydration. i've offered the patient admission for further ongoing hydration and trending of labs. However, she declines at this time and prefers discharge with close outpatient follow up.
    
    Additional history obtained from EMS.
    I have personally viewed and interpreted the labs, imaging, EKG and discussed all test results with the patient and updated her on the expected disease course.
    
    Disposition discharged to home.
    
    Diagnosis 
    fall from standing
    Facial contusion
    Dehydration.
    Traumatic rhabdomyolysis.
    Acute kidney injury
    
\end{lstlisting}

\subsection{Diagnosis and ICD-10-CM Code for Patient Note 3 - ER}
Fall from standing: W01.190A

Facial contusion: S00.83XA

Dehydration: E86.0

Traumatic rhabdomyolysis: T79.6XXA

Acute kidney injury: N17.9

\subsection{Patient Note 4 - ER}
\label{app:fictionalNote4}
\begin{lstlisting}
    Chief complaint.
    Jaw, clenching this morning, patient concern for seizure.
    
    HPI
    This is a 77-year-old female, the history of a prior stroke with prior seizure since into the emergency department after feeling her jaw clenching this morning. She thinks that it may have been seizure activity, however, was very different from prior seizures that she has experienced. She states she's never had jaw spasm like this before. She does have a dysarthria and aphasia from prior stroke at her baseline which she states unchanged. She reports that she's been compliant with her Keppra. She denies any recent infection or fever. Her husband states that she's had similar symptoms in the past and they were unable to find any diagnosis.
    
    Hospital medical history
    Seizure
    Stroke.
    
    Physical exam.
    Patient awake alert while appearing in no acute distress.
    Pupils are equal round reactive, extractive muscles are intact.
    Speech is with mild dysarthria, mild aphasia, patient reports at her baseline, strength and sensation are normal in all upper and lower extremities, face is symmetrical, oropharynx is clear.
    Normal professional, perfusion, regular rate and rhythm.
    Non-labored respirations with equal chest rise.
    
    Medical decision-making.
    Is the 77-year-old female with history of prior stroke and prior seizures presents into the emergency department after a transient spasm of her jaw
    Differential diagnosis includes muscle spasm, dehydration, occult infection, electrolyte, abnormalities, and cranial hemorrhage, low suspicion for focal seizure
    
    Patient has had no recurrence of her symptoms during the emergency department. EKG shows a sinus rhythm. Chemistry and CBC are reassuring. Component is negative. CT is negative for acute process. Your analysis is without signs of infection. I've offered the patient and her husband further observation, however they are more comfortable at this time with discharge. We will follow up with her neurologist as an outpatient. They are understanding of return precautions.
    
    Independent history obtained from EMS, patient's husband, patient's aid.
    I've independently viewed interpreted the labs, EKG, all imaging.
    
    Discussed all test results with patient and her husband And updated on expected disease course.
    
    Disposition discharged home.
    
    Diagnosis
    Facial spasm
    
\end{lstlisting}

\subsection{Diagnosis and ICD-10-CM Code for Patient Note 4 - ER}
Facial spasm: G51.39

\subsection{Patient Note 5 - ER}
\label{app:fictionalNote5}
\begin{lstlisting}
    CC: feeling anxious, short of breath, syncope

    HPI
    54 year old female patient with a history of asthma presents in to the ED for dyspnea. Patient states that she was at her physical therapist and began to feel short of breath so after her appointment she was going to her allergy immunologist. On the way there her e bike battery died and she had to walk the rest of the way. When she got there she felt very lightheaded and short of breath and then passed out. Transported here by EMS without intervention. Normal o2 on room air throughout. 
    
    Pmh
    Allergies
    Anxiety
    Asthma
    
    PE
    Awake and alert
    RRR no m/r/g, 2+dp,pt
    CTAB w/o w/r/r
    Neurointact nihss 0
    Abd sntnd 
    
    Mdm
    54 year old h/o asthma w/ syncope and dyspnea, resolved. 
    Differential includes anxiety, dehydration, arrhythmia, pe, ptx, pna
    
    Labs reassuring other than elevated d dimer. CTA obtained and negative. Patient feels improved with no recurrent symptoms. Ambulatory here with no lightheadedness. Offered further observation but declines and prefers discharge with outpatient follow up. 
    
    EKG on my interpretation with NSR narrow pr, narrow qrs, normal axis, no ischemic changes, normal qtc, no delta wave, no brugada, no epsilon wave. 
    
    I personally viewed and interpreted labs, imaging, ekg. 
    
    Disposition
    Discharge to home improved condition
    
    Diagnosis
    Syncope
    Dyspnea
    
\end{lstlisting}

\subsection{8.5.1	Diagnosis and ICD-10-CM Code for Patient Note 5 - ER}
Syncope: R55

Dyspnea: R06.00

\section{JSON Structures}
Structures presented here are verbatim copies of the JSON structures used in work through presentation in prompt context and in Ollama structured output parameters.
\subsection{Complex Output Structure}
\label{complexOutputStructure}
\begin{lstlisting}
"medical_record": {
    "original_document": "",
    "codes": {
        "diagnostic_codes": [],
        "procedure_codes": [],
        "billing_codes": []
    },
    "subjective": {
        "chief_complaint": "",
        "history_of_present_illness": "",
        "past_medical_history": "",
        "surgical_history": "",
        "pregnancy_history": "",
        "menstrual_history": "",
        "social_history": {
            "sexual_activity": "",
            "drug_use": "",
            "lifestyle": ""
        },
        "alcohol_use": "",
        "current_medications": [],
        "allergies": [],
        "review_of_systems": {
            "systems_reviewed": [
            {
                "system_name": "cardiovascular",
                "findings": ""
            },
            {
                "system_name": "respiratory",
                "findings": ""
            }
            ]
        }
    },
    "objective": {
        "vital_signs": {
            "temperature_celsius": "",
            "blood_pressure_mmHg": "",
            "heart_rate_bpm": "",
            "respiratory_rate_bpm": "",
            "oxygen_saturation_percent": ""
        },
        "physical_exam": [],
        "lab_results": [],
        "imaging": [],
        "diagnostic_procedures": []
    },
    "assessment": {
        "summary": "",
        "differential_diagnosis": [],
        "working_diagnosis": ""
    },
    "plan": {
        "expected_follow_up": "",
        "management_plan": [
            {
            "organ_system": "",
            "actions": []
            }
        ]
    },
    "orders": {
        "medications_ordered": [],
        "referrals_made": [],
        "labs_ordered": [],
        "imaging_ordered": []
    }
}
\end{lstlisting}

\subsection{Simple Output Structure}
\label{app:simpleOutputStructure}
\begin{lstlisting}
"medical_record": {
    "original_document": "",
    "codes": {
        "diagnostic_codes": [],
    },
    "subjective": {
        "chief_complaint": "",
    },
    "objective": {
        "vital_signs": {
            "temperature_celsius": "",
            "blood_pressure_mmHg": "",
            "heart_rate_bpm": "",
            "respiratory_rate_bpm": "",
            "oxygen_saturation_percent": ""
        },
        "physical_exam": "",
        "lab_results": [],
        "imaging": [],
        "diagnostic_procedures": [],
    },
    "assessment": {
        "summary": "",
        "differential_diagnosis": "",
        "working_diagnosis": "",
    },
    "plan": {
        "expected_follow_up": "",
        "management_plan": "",
    },
    "orders": {
        "referrals_made": [],
    }
}
\end{lstlisting}

\subsection{Trivial Output Structure}
\label{app:trivialOutputStructure}
\begin{lstlisting}
"medical_record": {
    "original_document": "",
    "diagnostic_codes": [],
    "diagnoses": [],
}

\end{lstlisting}

\section{Few-Shot Prompt}
\label{app:fewShotPrompt}
This is a verbatim reproduction of the few-shot prompt used in testing.
\begin{lstlisting}
    When provided with this doctors note:
    Chief complaint.
    Jaw, clenching this morning, patient concern for seizure.
    
    HPI
    This is a 77-year-old female, the history of a prior stroke with prior seizure since into the emergency department after feeling her jaw clenching this morning. She thinks that it may have been seizure activity, however, was very different from prior seizures that she has experienced. She states she's never had jaw spasm like this before. She does have a dysarthria and aphasia from prior stroke at her baseline which she states unchanged. She reports that she's been compliant with her Keppra. She denies any recent infection or fever. Her husband states that she's had similar symptoms in the past and they were unable to find any diagnosis.
    
    Hospital medical history
    Seizure
    Stroke.
    
    Physical exam.
    Patient awake alert while appearing in no acute distress.
    Pupils are equal round reactive, extractive muscles are intact.
    Speech is with mild dysarthria, mild aphasia, patient reports at her baseline, strength and sensation are normal in all upper and lower extremities, face is symmetrical, oropharynx is clear.
    Normal professional, perfusion, regular rate and rhythm.
    Non-labored respirations with equal chest rise.
    
    Medical decision-making.
    Is the 77-year-old female with history of prior stroke and prior seizures presents into the emergency department after a transient spasm of her jaw
    Differential diagnosis includes muscle spasm, dehydration, occult infection, electrolyte, abnormalities, and cranial hemorrhage, low suspicion for focal seizure
    
    Patient has had no recurrence of her symptoms during the emergency department. EKG shows a sinus rhythm. Chemistry and CBC are reassuring. Component is negative. CT is negative for acute process. Your analysis is without signs of infection. I've offered the patient and her husband further observation, however they are more comfortable at this time with discharge. We will follow up with her neurologist as an outpatient. They are understanding of return precautions.
    
    Independent history obtained from EMS, patient's husband, patient's aid.
    I've independently viewed interpreted the labs, EKG, all imaging.
    
    Discussed all test results with patient and her husband And updated on expected disease course.
    
    Disposition discharged home.
    
    Diagnosis
    Facial spasm
    
    The ouput should be:
    {
        "original_document": "Chief complaint.
            Jaw, clenching this morning, patient concern for seizure.
    
            HPI
            This is a 77-year-old female, the history of a prior stroke with prior seizure since into the emergency department after feeling her jaw clenching this morning. She thinks that it may have been seizure activity, however, was very different from prior seizures that she has experienced. She states she's never had jaw spasm like this before. She does have a dysarthria and aphasia from prior stroke at her baseline which she states unchanged. She reports that she's been compliant with her Keppra. She denies any recent infection or fever. Her husband states that she's had similar symptoms in the past and they were unable to find any diagnosis.
    
            Hospital medical history
            Seizure
            Stroke.
    
            Physical exam.
            Patient awake alert while appearing in no acute distress.
            Pupils are equal round reactive, extractive muscles are intact.
            Speech is with mild dysarthria, mild aphasia, patient reports at her baseline, strength and sensation are normal in all upper and lower extremities, face is symmetrical, oropharynx is clear.
            Normal professional, perfusion, regular rate and rhythm.
            Non-labored respirations with equal chest rise.
    
            Medical decision-making.
            Is the 77-year-old female with history of prior stroke and prior seizures presents into the emergency department after a transient spasm of her jaw
            Differential diagnosis includes muscle spasm, dehydration, occult infection, electrolyte, abnormalities, and cranial hemorrhage, low suspicion for focal seizure
    
            Patient has had no recurrence of her symptoms during the emergency department. EKG shows a sinus rhythm. Chemistry and CBC are reassuring. Component is negative. CT is negative for acute process. Your analysis is without signs of infection. I've offered the patient and her husband further observation, however they are more comfortable at this time with discharge. We will follow up with her neurologist as an outpatient. They are understanding of return precautions.
    
            Independent history obtained from EMS, patient's husband, patient's aid.
            I've independently viewed interpreted the labs, EKG, all imaging.
    
            Discussed all test results with patient and her husband And updated on expected disease course.
    
            Disposition discharged home.
    
            Diagnosis
            Facial spasm",
        "diagnostic_codes": ["G51.39"],
        "diagnoses": ["Facial spasm"]
    }
    
    When provided with a doctor's note like this:
    HPI:
    Is an 81 year-old female presenting into the emergency department after having a slip and fall at home. Patient states that she was on her way to the bathroom, she slipped on the bathroom floor and fell and hit her head. She says that she's been having back pain for the past one week and it's worsened at this time. The pain is constant, non-radiating. Pain worsens with any attempted movement. Denies any bladder incontinence. No loss of conscience. Not any anticoagulants. Denies any chest pain or shortness of breath.
    
    Past medical history:
    Denies any past medical history.
    Not on any medication's.
    
    Physical examination:
    Patient awake, alert, well appearing, no acute distress.
    Head is normal is normocephalic, atraumatic
    Normal peripheral perfusion with regular rate rhythm.
     Nonlabored respirations with equal chest rise, clear to auscultation 
    No focal CTL tenderness to palpation
    Focal tenderness to right greater trochanter and left mid tibia. NVI distal extremities. 
    
    medical decision-making:
    81-year-old presenting after a slip and fall.
    Differential diagnosis includes and cranial hemorrhage, contusion, fracture.
    
    Un reassessment patient remains hemodynamically stable and with non-focal neurologic exam. Plane films of the right hip and the left leg are negative for acute fracture, CT scan of the cervical spine, and the lumbar spine are negative for acute fracture or other acute traumatic injury. CT of the head shows a questionable asymmetric thickening of the inferior falx.
    
    Called to discuss with on-call neurosurgeon, in agreement with admission to our facility, repeat had CT at six hours. Patient and family are updated on all results and comfort a plan of care. Called and discussed with hospitalist for admission.
    
    I personally viewed an interpreted patient's EKG, chest, x-ray, plain films, CT imaging, laboratory results.
    
    Disposition - admit for further observation and repeat head CT
    
    Diagnosis
    Fall from standing
    Acute low back pain
    Abnormal imaging result
    
    The output should be:
    {
        "original_document": "HPI:
            Is an 81 year-old female presenting into the emergency department after having a slip and fall at home. Patient states that she was on her way to the bathroom, she slipped on the bathroom floor and fell and hit her head. She says that she's been having back pain for the past one week and it's worsened at this time. The pain is constant, non-radiating. Pain worsens with any attempted movement. Denies any bladder incontinence. No loss of conscience. Not any anticoagulants. Denies any chest pain or shortness of breath.
    
            Past medical history:
            Denies any past medical history.
            Not on any medication's.
    
            Physical examination:
            Patient awake, alert, well appearing, no acute distress.
            Head is normal is normocephalic, atraumatic
            Normal peripheral perfusion with regular rate rhythm.
            Nonlabored respirations with equal chest rise, clear to auscultation 
            No focal CTL tenderness to palpation
            Focal tenderness to right greater trochanter and left mid tibia. NVI distal extremities. 
    
            medical decision-making:
            81-year-old presenting after a slip and fall.
            Differential diagnosis includes and cranial hemorrhage, contusion, fracture.
    
            Un reassessment patient remains hemodynamically stable and with non-focal neurologic exam. Plane films of the right hip and the left leg are negative for acute fracture, CT scan of the cervical spine, and the lumbar spine are negative for acute fracture or other acute traumatic injury. CT of the head shows a questionable asymmetric thickening of the inferior falx.
    
            Called to discuss with on-call neurosurgeon, in agreement with admission to our facility, repeat had CT at six hours. Patient and family are updated on all results and comfort a plan of care. Called and discussed with hospitalist for admission.
    
            I personally viewed an interpreted patient's EKG, chest, x-ray, plain films, CT imaging, laboratory results.
    
            Disposition - admit for further observation and repeat head CT
    
            Diagnosis
            Fall from standing
            Acute low back pain
            Abnormal imaging result",
        "diagnostic_codes": [
            "W01.0XXA",
            "M54.50",
            "R93.0"
            ],
        "diagnoses": [
            "Fall from standing",
            "Acute low back pain",
            "Abnormal imaging result"
            ]
    }
    
\end{lstlisting}

\end{document}